\documentclass[lettersize,journal]{static/IEEEtran}
\usepackage{amsmath,amsfonts}
\usepackage{array}
\usepackage[caption=false,font=normalsize,labelfont=sf,textfont=sf]{subfig}
\usepackage{textcomp}
\usepackage{stfloats}
\usepackage{url}
\usepackage{verbatim}
\usepackage{cite}
\usepackage{algorithm} 
\usepackage{algpseudocode} 
\usepackage{graphicx,import}          
\usepackage{tikz}

\usepackage{mathrsfs} 

\usepackage{amssymb}
\usepackage[pdfpagelabels,  
bookmarks,       
pdftex
]{hyperref}
\hypersetup{
	pdfborder   = 0 0 0,
	plainpages  = false,
	bookmarksnumbered = true,
}
\usepackage{xcolor}
\usepackage[nolist, nohyperlinks, printonlyused, withpage, smaller]{acronym}
\newtheorem{Theorem}{Theorem}
\newtheorem{Lemma}{Lemma}
\newtheorem{Definition}{Definition}
\newtheorem{Assumption}{Assumption}
\newtheorem{Remark}{Remark}

\newcommand\copyrighttext{%
	\footnotesize This work has been submitted to the IEEE for possible publication. Copyright may be transferred without notice, after which this version may no longer be accessible.}

\newcommand\copyrightnotice{%
	\begin{tikzpicture}[remember picture,overlay]
		\node[anchor=north,yshift=-5mm] at (current page.north) {\fbox{\parbox{\dimexpr\textwidth-\fboxsep-\fboxrule\relax}{\copyrighttext}}};
	\end{tikzpicture}%
}

\begin{document}

\begin{acronym}
    \acro{mbilc}[MB-ILC]{Model-Based Iterative Learning Control}
    \acro{fdilc}[FD-ILC]{Frequency-Domain Iterative Learning Control}
	\acro{rnn}[RNN]{Recurrent Neural Network}
	\acroplural{rnn}[RNN]{Recurrent Neural Networks}
	\acro{narx}[NARX-NN]{Nonlinear Autoregressive Exogenous Neural Network}
	\acroplural{narx}[NARX-NN]{Nonlinear Autoregressive Exogenous Neural Networks}
	\acro{gru}[GRU]{{Gated Recurrent Unit}}
	\acroplural{gru}[GRU]{{Gated Recurrent Units}}
	\acro{lstm}[LSTM]{Long Short-Term Memory}
	\acro{ann}[NN]{Neural Network}	
	\acroplural{ann}[NN]{Neural Networks}
	\acro{ffnn}[FFNN]{Feedforward Neural Network}
	\acroplural{ffnn}[FFNN]{Feedforward Neural Networks}
	\acro{pinn}[PINN]{Physics-Informed Neural Network}
	\acroplural{pinn}[PINN]{Physics-Informed Neural Networks}
	\acro{gp}[GP]{Gaussian Process}
	\acroplural{gp}[GP]{Gaussian Processes}
	\acro{knn}[$K$NN]{$K$-Nearest-Neighbors}
	\acro{ilc}[ILC]{Iterative Learning Control}
	\acro{ili}[ILI]{Iterative Learning Identification}
	\acro{rc}[RC]{Repetitive Control}
	\acro{rl}[RL]{{Reinforcement Learning}}
	\acro{daoc}[DAOC]{{Direct Adaptive Optimal Control}}
	\acro{ml}[ML]{maschinelles Lernen}
	\acro{lwpr}[LWPR]{{Locally Weighted Projection Regression}}
	\acro{svm}[SVM]{{Support Vector Machine}}
	\acro{mcmc}[MCMC]{{Markov Chain Monte Carlo}}
	\acro{ad}[AD]{{Automatic Differentiation}}
	\acro{gmm}[GMM]{{Gaussian Mixture Models}}
	\acro{rkhs}[RKHS]{{Reproducing Kernel Hilbert Spaces}}
	\acro{rbf}[RBF]{{Radial Basis Function}}
	\acro{rbfnn}[RBF-NN]{{Radial Basis Function Neural Network}}
	\acroplural{rbfnn}[RBF-NN]{{Radial Basis Function Neural Networks}}
	\acro{node}[{Neural} ODE]{{Neural Ordinary Differential Equation}}
	\acroplural{node}[{Neural} ODEs]{{Neural Ordinary Differential Equations}}
	\acro{pdf}[PDF]{Probability Density Function}
	\acro{pca}[PCA]{Principal Component Analysis}

	\acro{kf}[KF]{Kalman Filter}	
	\acro{ekf}[EKF]{Extended Kalman Filter}
	\acro{nekf}[NEKF]{Neural Extended Kalman Filter}
	\acro{ukf}[UKF]{Unscented Kalman Filter}
	\acro{pf}[PF]{Particle Filter}
	\acro{mhe}[MHE]{{Moving Horizon Estimation}}
	\acro{rpe}[RPE]{{Recursive Predictive Error}}
	\acro{rls}[RLS]{{Recursive Least Squares}}
	\acro{slam}[SLAM]{{Simultaneous Location and Mapping}}
	
	\acro{mftm}[MFTM]{Magic Formula Tire Model}
	\acro{mbs}[MBS]{Multi-Body Simulation}
	\acro{lti}[LTI]{Linear Time-Invariant}
	\acro{cog}[COG]{Center Of Gravity}
	\acro{ltv}[LTV]{Linear Time-Variant}
	\acro{siso}[SISO]{Single-Input Single-Output}
	\acro{mimo}[MIMO]{Multiple-Input Multiple-Output}
	\acro{psd}[PSD]{Power Spectral Density}
	\acroplural{psd}[PSD]{Power Spectral Densities}
	\acro{cf}[CF]{Coordinate Frame}
	\acroplural{cf}[CF]{Coordinate Frames}

	\acro{pso}[PSO]{Particle Swarm Optimization}
	\acro{sqp}[SQP]{Sequentielle Quadratische Programmierung}
	\acro{svd}[SVD]{Singular Value Decomposition}
	\acro{ode}[ODE]{Ordinary Differential Equation}
	\acroplural{ode}[ODE]{Ordinary Differential Equations}
	\acro{pde}[PDE]{Partial Differential Equation}
	
	\acro{nmse}[NMSE]{Normalized Mean Squared Error}
	\acro{mse}[MSE]{Mean Squared Error}
	\acro{rmse}[RMSE]{Root Mean Squared Error}
	\acro{wrmse}[wRMSE]{weighted \ac{rmse}}
	
	\acro{mpc}[MPC]{{Model Predictive Control}}
	\acro{nmpc}[NMPC]{Nonlinear Model Predictive Control}
	\acro{lmpc}[LMPC]{Learning Model Predictive Control}
	\acro{ltvmpc}[LTV-MPC]{\acl{ltv} Model Predictive Control}
	\acro{ndi}[NDI]{Nonlinear Dynamic Inversion}
	\acro{ac}[AC]{Adhesion Control}
	\acro{esc}[ESC]{Electronic Stability Control}
	\acro{ass}[ASS]{Active Suspension System}
	\acro{trc}[TRC]{Traction Control}
	\acro{abs}[ABS]{Anti-Lock Brake System}
	\acro{gcc}[GCC]{Global Chassis Control}
	\acro{ebs}[EBS]{Electronic Braking System}
	\acro{adas}[ADAS]{Advanced Driver Assistance Systems}
	\acro{cm}[CM]{{Condition Monitoring}}
	\acro{hil}[HiL]{{Hardware-in-the-Loop}}

	\acro{irw}[IRW]{Independently Rotating Wheels}
	\acro{dirw}[DIRW]{Driven \acl{irw}}
	\acro{imes}[imes]{{Institute of Mechatronic Systems}}
	\acro{db}[DB]{Deutsche Bahn}
	\acro{ice}[ICE]{Intercity-Express}
	
	\acro{fmi}[FMI]{{Functional Mock-up Interface}}
	\acro{fmu}[FMU]{{Functional Mock-up Unit}}
	\acro{doi}[DOI]{{Digital Object Identifier}}

	\acro{ra}[RA]{Research Area}
	\acroplural{ra}[RA]{Research Areas}
	\acro{wp}[WP]{Work Package}
	\acroplural{wp}[WP]{Work Packages}
	
	\acro{fb}[FB]{Forschungsbereich}
	\acroplural{fb}[FB]{Forschungsbereiche}
	\acro{ap}[AP]{Arbeitspaket}
	\acroplural{ap}[AP]{Arbeitspakete}
	\acro{abb}[Abb.]{Abbildung}
	\acro{luis}[LUIS]{Leibniz Universit"at IT Services}
	
	\acro{res}[RES]{Renewable Energy Sources}
	\acro{pkw}[PKW]{Personenkraftwagen}
	\acro{twipr}[TWIPR]{{Three-Wheeled Inverted Pendulum Robot}}
	\acro{tlr}[TLR]{Two-Link Robot}
	
	\acro{pmcmc}[PMCMC]{Particle Markov Chain Monte Carlo}
	\acro{mcmc}[MCMC]{Markov Chain Monte Carlo}
	\acro{rbpf}[RBPF]{Rao-Blackwellized Particle Filter}
	\acro{pf}[PF]{Particle Filter}
	\acro{ps}[PS]{Particle Smoother}
	\acro{smc}[SMC]{Sequential Monte Carlo}
	\acro{csmc}[cSMC]{conditional SMC}
	\acro{mh}[MH]{Metropolis Hastings}
	\acro{em}[EM]{Expectation Maximization}
	\acro{slam}[SLAM]{Simultaneous Location and Mapping}
	\acro{dof}[DOF]{Degree of Freedom}
	\acroplural{dof}[DOF]{Degrees of Freedom}
	\acro{pg}[PG]{Particle Gibbs}
	\acro{pgas}[PGAS]{Particle Gibbs with Ancestor Sampling}
	\acro{mpgas}[mPGAS]{marginalized Particle Gibbs with Ancestor Sampling}
	\acro{hmm}[HMM]{Hidden Markov Model}
	
	\acro{emps}[EMPS]{Electro-Mechanical Positioning System}
	
	\acro{ilc}[ILC]{Iterative Learning Control}
	\acro{ddilc}[DD-ILC]{Data-Driven Iterative Learning Control}
	\acro{dilc}[DILC]{Dual Iterative Learning Control}
	\acro{iml}[IML]{Iterative Model Learning}
	\acro{noilc}[NO-ILC]{Norm-Optimal Iterative Learning Control}
	\acro{gilc}[G-ILC]{Gradient Iterative Learning Control}
	\acro{noiml}[NO-IML]{Norm-Optimal Iterative Model Learning}
	\acro{giml}[G-IML]{Gradient Iterative Model Learning}

	\acro{ggdilc}[GG-DILC]{GG-Dual Iterative Learning Control}
	\acro{gnodilc}[GNO-DILC]{GNO-Dual Iterative Learning Control}
    \acro{nogdilc}[NOG-DILC]{NOG-Dual Iterative Learning Control}
    \acro{nonodilc}[NONO-DILC]{NONO-Dual Iterative Learning Control}

\end{acronym}

\newcommand{\longcomment}[1]{}

\newcommand{\qvec}[1]{\mathbf{#1}}
\newcommand{\qmat}[1]{\mathbf{#1}}
\newcommand{\idx}[1]{_{\mathrm{#1}}}

\newcommand{\qRealNumbers}{\mathbb{R}}
\newcommand{\qPositiveRealNumbers}{\mathbb{R}_{\geq 0}}
\newcommand{\qNaturalNumbersZero}{\mathbb{N}_{\geq 0}}
\newcommand{\qNaturalNumbersPos}{\mathbb{N}_{>0}}
\newcommand{\qNaturalNumbers}{\mathbb{N}}
\newcommand{\foralljinN}{\forall j \in \qNaturalNumbersZero, \quad}
\newcommand{\foralljinNPos}{\forall j \in \qNaturalNumbersPos, \quad}
\newcommand{\qBLT}{\mathcal{T}^{\mathrm{BLT}}}

\newcommand{\qa}{\qvec{a}}
\newcommand{\qb}{\qvec{b}}
\newcommand{\qc}{\qvec{c}}
\newcommand{\qd}{\qvec{d}}
\newcommand{\qe}{\qvec{e}}
\newcommand{\qf}{\qvec{f}}
\newcommand{\qg}{\qvec{g}}
\newcommand{\qh}{\qvec{h}}
\newcommand{\qi}{\qvec{i}}
\newcommand{\qj}{\qvec{j}}
\newcommand{\qk}{\qvec{k}}
\newcommand{\ql}{\qvec{l}}
\newcommand{\qm}{\qvec{m}}
\newcommand{\qn}{\qvec{n}}
\newcommand{\qo}{\qvec{o}}
\newcommand{\qp}{\qvec{p}}
\newcommand{\qq}{\qvec{q}}
\newcommand{\qr}{\qvec{r}}
\newcommand{\qs}{\qvec{s}}
\newcommand{\qt}{\qvec{t}}
\newcommand{\qu}{\qvec{u}}
\newcommand{\qv}{\qvec{v}}
\newcommand{\qw}{\qvec{w}}
\newcommand{\qx}{\qvec{x}}
\newcommand{\qy}{\qvec{y}}
\newcommand{\qz}{\qvec{z}}
\newcommand{\qem}{\qe^{\mathrm{m}}}

\newcommand{\qA}{\qvec{A}}
\newcommand{\qB}{\qvec{B}}
\newcommand{\qC}{\qvec{C}}
\newcommand{\qD}{\qvec{D}}
\newcommand{\qE}{\qvec{E}}
\newcommand{\qF}{\qvec{F}}
\newcommand{\qG}{\qvec{G}}
\newcommand{\qH}{\qvec{H}}
\newcommand{\qI}{\qvec{I}}
\newcommand{\qJ}{\qvec{J}}
\newcommand{\qK}{\qvec{K}}
\newcommand{\qL}{\qvec{L}}
\newcommand{\qM}{\qvec{M}}
\newcommand{\qN}{\qvec{N}}
\newcommand{\qO}{\qvec{O}}
\newcommand{\qP}{\qvec{P}}
\newcommand{\qQ}{\qvec{Q}}
\newcommand{\qR}{\qvec{R}}
\newcommand{\qS}{\qvec{S}}
\newcommand{\qT}{\qvec{T}}
\newcommand{\qU}{\qvec{U}}
\newcommand{\qV}{\qvec{V}}
\newcommand{\qW}{\qvec{W}}
\newcommand{\qX}{\qvec{X}}
\newcommand{\qY}{\qvec{Y}}
\newcommand{\qZ}{\qvec{Z}}

\newcommand{\quv}{\bar{\qu}}
\newcommand{\qyv}{\bar{\qy}}
\newcommand{\qxv}{\bar{\qx}}
\newcommand{\qZero}{\qvec{0}}
\newcommand{\qdu}{\boldsymbol{\Delta}\qu}
\newcommand{\qdy}{\boldsymbol{\Delta}\qy}
\newcommand{\qep}{\hat{\qe}}
\newcommand{\qyp}{\hat{\qy}}

\newcommand{\qILCDesign}{\boldsymbol{D}}
\newcommand{\qIMLDesign}{\boldsymbol{\hat{D}}}
\newcommand{\qLIML}{\hat{\qL}}
\newcommand{\qeR}{\qe_\mathrm{R}}

\newcommand{\grad}{\boldsymbol{\nabla}}

\newcommand{\qff}{\boldsymbol{f}}
\newcommand{\qpf}{\boldsymbol{p}}
\newcommand{\qmf}{\boldsymbol{m}}
\newcommand{\qXf}{\boldsymbol{A}}

\newcommand{\qLift}{\boldsymbol{\mathcal{L}}}
\newcommand{\qLu}{\boldsymbol{\mathcal{L}}\idx{u}}
\newcommand{\qLm}{\boldsymbol{\mathcal{L}}\idx{m}}

\newcommand{\qUb}{\bar{\qU}}
\newcommand{\qTb}{\bar{\qT}}
\newcommand{\qVb}{\bar{\qV}}
\newcommand{\qyh}{\hat{\qy}}
\newcommand{\qyt}{\tilde{\qy}}
\newcommand{\qeh}{\hat{\qe}}
\newcommand{\qLh}{\hat{\qL}}
\newcommand{\qDe}{{\boldsymbol{\mathcal{D}}}}
\newcommand{\qDeh}{\hat{\boldsymbol{\mathcal{D}}}}

\newcommand{\qTwoNorm}[1]{\left\| {#1} \right\|_{2}}
\newcommand{\qInfNorm}[1]{\left\| {#1} \right\|_\infty}
\newcommand{\qOneNorm}[1]{\left\| {#1} \right\|\idx{1}}
\newcommand{\qNorm}[1]{\left\| {#1} \right\|}
\newcommand{\qGivenNorm}{\qNorm{\boldsymbol{\cdot}}}

\newcommand{\qred}[1]{{\color{red}#1}}
\newcommand{\imesorange}{E77B29}
\newcommand{\imesgruen}{C8D317}
\newcommand{\imesblauHundert}{00509B}
\newcommand{\imesblauZwanzig}{CCDCEB}
\newcommand{\imesblauVierzig}{99B9D8}

\newcommand{\Lfunc}{\mathscr{L}}       
\newcommand{\Kfunc}{\mathscr{K}}
\newcommand{\KLfunc}{\Kfunc\negthinspace\negthinspace\Lfunc}

\newcommand{\eg}{e.\,g.,\,}
\newcommand{\ie}{i.\,e.,\,}
\newcommand*{\R}{\mathbb{R}}

\newcommand*{\MNIW}{\mathcal{MNIW}}
\newcommand*{\IW}{\mathcal{IW}}
\newcommand*{\T}{\mathcal{T}}
\newcommand*{\N}{\mathcal{N}}

\newcommand{\del}{\partial}
\newcommand{\bi}[1]{\boldsymbol{#1}}
\newcommand{\ur}[1]{\mathrm{#1}}
\newcommand{\cali}[1]{\mathcal{#1}}
\newcommand\norm[1]{\left\lVert#1\right\rVert}

\newcommand{\ubar}[1]{\underaccent{\bar}{#1}}
\newcommand{\ToDo}[1]{\todo[size=\tiny]{#1}}

\newcommand{\spans}[1]{\spanop\left( #1 \right)}

\newcommand{\ToDos}[1]{\todo[size=\tiny]{#1}}


\title{Dual Iterative Learning Control for Multiple-Input Multiple-Output Dynamics with Validation in Robotic Systems}

\author{Jan-Hendrik Ewering, Alessandro Papa, Simon F.\,G. Ehlers, Thomas Seel, and Michael Meindl
\thanks{The authors are with the Institute of Mechatronic Systems, Leibniz Universität Hannover, Garbsen, Germany.}
\thanks{This work was supported by the German Academic Scholarship Foundation (Studienstiftung des Deutschen
Volkes).}
\thanks{Manuscript received September 23, 2025; revised XX, 2025.}}



\maketitle\copyrightnotice\vspace{-9.5pt}
\thispagestyle{empty}

\begin{abstract}
Solving motion tasks autonomously and accurately is a core ability for intelligent real-world systems.
To achieve genuine autonomy across multiple systems and tasks, key challenges include coping with unknown dynamics and overcoming the need for manual parameter tuning, which is especially crucial in complex \ac{mimo} systems.

This paper presents \ac{mimo} \ac{dilc}, a novel data-driven iterative learning scheme for simultaneous tracking control and model learning, without requiring any prior system knowledge or manual parameter tuning. 
The method is designed for repetitive \ac{mimo} systems and integrates seamlessly with established iterative learning control methods. 
We provide monotonic convergence conditions for both reference tracking error and model error in linear time-invariant systems.

The \ac{dilc} scheme---rapidly and autonomously---solves various motion tasks in high-fidelity simulations of an industrial robot and in multiple nonlinear real-world \ac{mimo} systems, without requiring model knowledge or manually tuning the algorithm. 
In our experiments, many reference tracking tasks are solved within 10-20 trials, and even complex motions are learned in less than 100 iterations. 
We believe that, because of its rapid and \textit{autonomous} learning capabilities, \ac{dilc} has the potential to serve as an efficient building block within complex learning frameworks for intelligent real-world systems.
\end{abstract}


\section{Introduction}
\label{sec1}
\acresetall
\IEEEPARstart{A}{ccurate} reference tracking is a critical control capability for a wide range of real-world applications, from industrial manufacturing to service robotics and biomedical systems \cite{Michalos.2016, Murphy.2004, Wang.2011}, which often involve complex \ac{mimo} system dynamics. 
For these systems to be effective and user-friendly, they must be capable of self-reliantly adapting to new tasks and environments. In other words, it is required to learn to perform reference tracking autonomously. 
This capability is crucial to eliminate the need for expert-provided model information or time-consuming manual tuning.

In repetitive settings, \ac{ilc} is an established method that enables highly accurate reference tracking, given a reference trajectory \cite{Arimoto.1984, Bristow.2006, Wang.2009}. 
However, satisfactory learning performance is almost always dependent on human expert knowledge, such as access to model information or the manual tuning of algorithmic (hyper) parameters. 
The necessary manual tuning effort is typically even more aggravated in \ac{mimo} systems due to cross-coupling effects and scale variations between different inputs and outputs \cite{Freeman.2015}. 
Moreover, learning performance is often system- or reference-specific, which hinders genuine autonomous deployment across numerous systems and tasks. 

Hence, an \ac{ilc} method must possess the following three characteristics in order to enable real-world systems to autonomously learn to solve reference tracking tasks. 
First, the \ac{ilc} method must neither require prior model information nor the manual tuning of parameters to enable autonomous application.
Second, the \ac{ilc} method has to be applicable to \ac{mimo} dynamics, as these are often present in realistic settings. 
And third, the \ac{ilc} method should be validated -- ideally on multiple -- real-world systems.
Based on these criteria, we continue to review the state of research in \ac{ilc}.

First, there exists an extensive class of so-called \ac{mbilc} methods that can yield remarkable reference tracking performance in various real-world applications.
For example, \ac{noilc} schemes have been applied to gantry robots \cite{Ratcliffe.2006} and stroke rehabilitation \cite{Rogers.2010}, and \ac{noilc} can readily be applied to \ac{mimo} dynamics \cite{Owens.2013,Owens.2016}.
Similarly, \ac{fdilc} has been successfully applied to real-world systems with \ac{mimo} dynamics such as a marine vibrator \cite{Sornmo.2016} or nano-positioning systems \cite{Helfrich.2010}. 
Despite these achievements, the aforementioned and other \ac{mbilc} methods are limited because they require prior model information and typically involve manual tuning of learning parameters, which hinders their autonomous application.

To overcome the need for prior model information, so-called \ac{ddilc} methods have been developed. 
A common approach is to use the input/output trajectory pairs from previous trials to estimate the gradient of the tracking errors with respect to the input trajectory to update the latter \cite{Bolder.2018, Aarnoudse.2021, Huo.2019}. 
On the other hand, several \ac{ddilc} schemes use experimental data to estimate a model of the plant dynamics and combine this plant approximation with well-known \ac{mbilc} methods \cite{Owens.2016, Chu.2025}.  
For instance, some approaches use \ac{rls} to estimate a model of the system dynamics and combine it with \ac{noilc} or adaptive \ac{ilc} \cite{Janssens.2013, Chai.2023, Tzou.1999}. 
Other approaches combine \ac{ili} and \ac{ilc} \cite{Liu.2016b, Hou.2015} to iteratively learn a model and input trajectory without prior model information.
\ac{fdilc} has been combined with an iterative learning approach for \ac{mimo} dynamics using a pseudo-inversion approach \cite{Rozario.2019}. 
Notably, there are approaches that iteratively learn a dynamic linearization of the plant dynamics, which can be utilized in a \ac{noilc} update law \cite{Chi.2012, Chi.2015, Chi.2017, Hou.2017, Yu.2020b}.
What is common among all of these methods is that they overcome the need for prior model information, and many of them are applicable to \ac{mimo} dynamics. 
However, most of these methods have \emph{not} been validated in real-world experiments, and all of these methods require the manual tuning of learning parameters, which precludes the autonomous application of the learning methods. 

To overcome the need for prior model information \textit{and} manual parameter tuning, \ac{mbilc} has been combined with repeated model learning using Gaussian processes and self-parametrization schemes \cite{Meindl.2024,Meindl.2022,Meindl.2024c}, and some of the methods have been validated on different real-world systems \cite{Meindl.2024}. 
However, these approaches are limited in terms of their applicability and validation in \ac{mimo} dynamics. 
We, hence, conclude that there is no \ac{ddilc} method that is autonomous in the sense that it neither requires prior model information nor manual parameter tuning, is applicable to \ac{mimo} dynamics, and has been validated on multiple---possibly real-world---systems.

\longcomment{
\subsection{Old Version}
Existing \ac{ddilc} approaches can be broadly categorized into model-based \ac{ddilc} \cite{Janssens.2013} and model-free \ac{ddilc} \cite{Aarnoudse.2021}. 

On the one hand, \textbf{model-free \ac{ddilc}} does not rely on a (potentially approximated) system model for control learning. Being independent of assumptions about the underlying system structure, model-free \ac{ddilc} schemes are often flexibly applicable to a wide range of scenarios. Recent approaches rely on, for instance, gradient approximations obtained from input-output data of the system, either by probing the system between successive iterations \cite{Bolder.2018} or by considering successive trials \cite{Aarnoudse.2021}.
\qred{model-free \ac{ddilc}
\begin{itemize}
	\item \cite{Aarnoudse.2021} presents model-free \ac{ilc} for \ac{mimo} systems based on a gradient estimate. The gradient estimate requires a single experiment at each iteration and relies on a stochastic gradient descent algorithm. However, no real-world validation is conducted. Moreover, the paper does not exploit the duality of \ac{iml} and \ac{ilc}. 
	\item \cite{Huo.2019} presents a gradient-based model-free \ac{ilc} scheme for nonlinear \ac{mimo} systems. It relies on a gradient estimate obtained from probing the system between successive learning steps. However, no real-world validation is conducted. Moreover, the paper does not exploit the duality of \ac{iml} and \ac{ilc} approaches.
\end{itemize}}
\qred{Structure thought: model-free \ac{ddilc} schemes often rely on many user design choices and hyperparameters to be tuned, which is to be expected, as no further structure, such as model information, guides the control learning procedure (Maybe provide concrete examples that support this point). This hampers the ability of model-free \ac{ddilc} to solve different reference tracking tasks in various systems truly autonomously without (re) parametrization of an expert user.}
Despite their flexibility, these approaches typically require extensive parameter tuning and many learning trials to achieve satisfactory performance, a limitation that is particularly severe in \ac{mimo} systems where cross-coupling effects and increasing input- and output dimensions significantly complicate the learning process. 
Moreover, many state-of-research research approaches are not validated in real-world experiments with varying tasks and different system dynamics, leaving the question of truly autonomous learning open.

\IEEEpubidadjcol
On the other hand, \textbf{model-based \ac{ddilc}} methods typically approximate a system model from measured input-output data and enable seamless integration of well-known \ac{noilc} ideas \cite{Owens.2016, Chu.2025}. Moreover, an approximately known system model facilitates, in principle, the generalization to trial-varying references. In \cite{Janssens.2013}, a model-based \ac{ddilc} ... 
\qred{model-based \ac{ddilc}
\begin{itemize}
	\item  Presents an adaptive \ac{ilc} framework that executes control and model learning consecutively and does not require model information. It uses \ac{rls} for adaptive model learning. However, it does not establish a duality between \ac{ilc} and \ac{iml}, and it considers only \ac{siso} systems. Moreover, it does not provide a validation on real-world systems.
	\item \cite{Hou.2015} does \ac{ili} combined with \ac{ilc} for joint model and control input learning. However, it does not provide \ac{mimo} results. Moreover, it does not provide a validation on real-world systems.
	\item \cite{Chai.2023} does \ac{ilc} with a model that is learned recursively by \ac{rls}. No duality between \ac{ilc} and \ac{rls}/\ac{iml} is exploited. The paper considers only \ac{siso} systems, and does not provide a thorough validation on real-world systems.
	\item \cite{Tzou.1999} shows an early data-driven \ac{ilc} scheme that employs \ac{rls} for iterative learning of the system model and using it within adaptive \ac{ilc}. It does not exploit the duality of \ac{iml} and \ac{ilc}. However, it considers only \ac{siso} systems, and does not provide a thorough validation on multiple real-world systems.
	\item \cite{Liu.2016b} presents an integrated \ac{ili} and \ac{ilc} scheme with joint convergence conditions, but does not exploit and investigate self-parametrization. Moreover, it considers only \ac{siso} systems and does not provide a thorough validation on multiple real-world systems.
	\item \cite{Rozario.2019} present a frequency-domain data-driven \ac{ilc} scheme for \ac{mimo} systems. It involves an iterative model learning mechanism that relies on pseudo-inversion of the plant from trial to trial. A smoothing mechanism to stabilize the pseudo-inversion is presented. However, the paper does not exploit the duality of \ac{iml} and \ac{ilc}. 
	\item \cite{Janssens.2013} presents a framework for data-driven \ac{noilc} that identifies the Markov parameters of a system based on a linear combination of past inputs and outputs. They elaborate on how an uncertainty estimate of the Markov parameters can be used for parametrization of the \ac{noilc} scheme, trading off learning robustness and trial efficiency. However, the paper focuses on \ac{siso} systems and lacks a representative validation on multiple real-world systems. Moreover, the paper does not establish a duality of \ac{iml} and \ac{ilc} approaches.
	\item \cite{Meindl.2024,Meindl.2022,Meindl.2024c} implement data-driven \ac{ilc} schemes for dynamical systems that enable truly autonomous learning without model information or (re-)tuning on a wide range of systems, validated in several real-world experiments. However, they do not exploit the duality of \ac{iml} and \ac{ilc} approaches, and do not provide conditions for convergence and stability. In addition, they do not provide \ac{mimo} validation results.
	\item \cite{Meindl.2025} presents a novel iterative learning paradigm that enables to learn models through the lens of \ac{ilc} approaches. However, it does not provide results for \ac{mimo} systems: no theoretical results, no \ac{mimo} self-parametrization, no \ac{mimo} validation. Although the class of \ac{mimo} systems is most important for real-world applicability, and control in these systems comes with significant challenges, such as coupled dynamics and scale variations between different inputs and outputs.
\end{itemize}}
\qred{Structure thought: In model-based \ac{ddilc}, the model information provides a lot of structure that avoids massive hyperparameter tuning and enables self-tuning capabilities. In this light, model-based \ac{ddilc} frameworks represent efficient policy design laws, in the sense that few further design choices beyond the model knowledge itself need to be taken.}
However, the vast majority of existing model-based \ac{ddilc} methods lack a consistent scheme for \ac{iml} with performance guarantees. While some works provide a solid basis for such results \cite{Meindl.2025, Liu.2016}, they remain limited to \ac{siso} systems and lack a thorough real-world validation with multiple systems and different tasks.
}

\begin{figure*}[t]
	\centering
	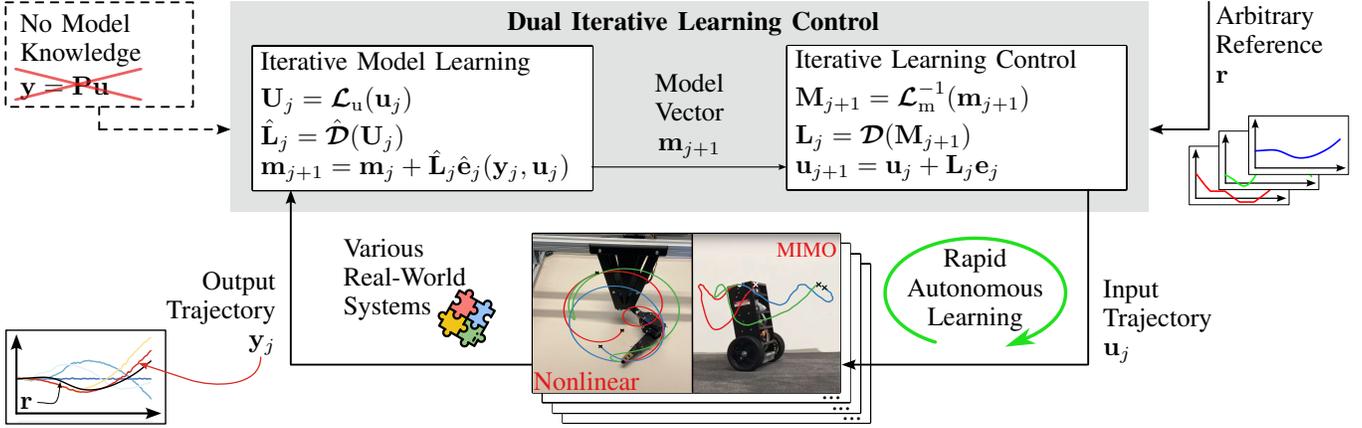
	\caption{Overview of the \acl{dilc} (\acs{dilc}) scheme for autonomous motion learning without any model information or manual (re-)tuning. Given an arbitrary reference $\qr$, three steps are executed iteratively, trial by trial, for simultaneously learning both the control input and the system model. First, an experiment is conducted to obtain input-output data $(\qu_j, \qy_j)$. Second, the model parameters are updated by \acl{iml} (\acs{iml}) using established \ac{ilc} principles, which is enabled by the lifting operations $\qLu$ and $\qLm$. Third, the control input is learned using the updated model within model-based \ac{ilc}.}
	\label{fig:architecture}
\end{figure*}

To address these three issues, we propose a novel \ac{mimo} \ac{dilc} framework that builds on previous results \cite{Meindl.2025} and enables autonomous learning of reference tracking tasks in real-world systems with \ac{mimo} dynamics. 
Specifically, the contributions of this paper are threefold:
\begin{itemize}
	\item First, a \textbf{novel \ac{dilc}} scheme for simultaneous model and control learning in \ac{mimo} systems, while requiring neither prior model information nor manual parameter tuning. It exploits a novel iterative learning paradigm that generalizes \ac{ilc} approaches for iterative model learning, thus enabling the learning of system models using established \ac{ilc} methods. The algorithmic architecture is illustrated in Figure \ref{fig:architecture}.
	\item Second, a \textbf{theoretical analysis} providing convergence conditions of the proposed algorithm under mild assumptions. We emphasize that iterative model learning in complex \ac{mimo} systems poses significant challenges, such as an overparametrized model, for which we present novel analysis to prove convergence. 
	\item Third, an extensive \textbf{empirical validation} with two real-world \ac{mimo} systems and a six-degree-of-freedom industrial robot simulation. We demonstrate, in contrast to the vast majority of existing works, the truly autonomous learning capabilities of \ac{dilc} without any model information or human tuning effort. To the best of our knowledge, this is the first time that a \ac{ddilc} method has solved different reference tracking tasks in multiple real-world systems with \ac{mimo} dynamics, without requiring prior model information or manual parameter tuning. We highlight that \ac{dilc} solves many reference tracking tasks within $10$-$20$ trials and learns even complex motions in less than $100$ iterations.
\end{itemize}

\longcomment{
To overcome this issue, we build on the results of \cite{Meindl.2025}, which introduced the concept of \ac{dilc} for \ac{siso} dynamics.
In this paper,\ac{dilc} is extended to be applicable to systems with \ac{mimo} dynamics.
\ac{dilc} iteratively learns both a system model and a control input trajectory, while requiring neither prior model information nor manual parameter tuning. 
To the best of our knowledge, this is the first method that \ac{ddilc} methods solve different reference tracking tasks in multiple real-world systems with \ac{mimo} dynamics without requiring prior model information or manual parameter tuning.
We highlight that it learns within a very small number of trials and in complex \ac{mimo} systems, including a six-degree-of-freedom industrial robot simulation. 
Specifically, the contributions of this paper are threefold:
\begin{itemize}
	\item First, a \textbf{novel \ac{dilc}} scheme for simultaneous model and control learning in \ac{mimo} systems without human expert information. It exploits a novel iterative learning paradigm that generalizes \ac{ilc} approaches for iterative model learning and thus enables learning system models using established \ac{ilc} methods.
	\item Second, a \textbf{theoretical analysis} proving convergence conditions of the proposed algorithm under mild conditions.
	\item Third, an extensive \textbf{empirical validation}, demonstrating, in contrast to the vast majority of existing works, the truly autonomous learning capabilities of \ac{dilc} without any model information or human tuning effort.
\end{itemize}}

This paper is structured as follows. 
We formally define the considered problem in Section \ref{sec:problem} and introduce preliminaries on \ac{ilc} in Section \ref{sec:preliminaries}. 
The proposed method and its theoretical properties are detailed in Section \ref{sec:methods}. 
The simulative and experimental results are presented in Section \ref{sec:results}. 
Finally, we conclude the paper in Section \ref{sec:conclusion}.\\

\emph{Notation:} We denote the set of real numbers by $\qRealNumbers$, the set of natural numbers by $\qNaturalNumbers$, the set of all natural numbers greater than or equal to $a \in \qNaturalNumbers$ by $\qNaturalNumbers_{\geq a}$, and the set of natural numbers in the interval $[a, b] \subset \qNaturalNumbers$ by $\qNaturalNumbers_{[a,b]}$. 
We denote vectors (matrices) by lower-case (upper-case) letters in bold, \eg $\qv \in \qRealNumbers^{N}$ ($\qA\in \qRealNumbers^{N \times N}$). 
If not explicitly stated, all vectors are column vectors, and by writing $[\qv]_{i}$, we refer to the $i$-th entry of $\qv$. 
By writing $[\qA]_{i,j}$, we refer to the $i$-th entry of the $j$-th column of $\qA$. 
To vectorize $\qA$, we write $\mathrm{vec}(\qA)$. 
The Euclidean norm of a vector $\qv$ is denoted by $\norm{\qv}$, and the induced Euclidean norm of a matrix $\qA$ is denoted by $\norm{\qA}$. 
The weighted norm with respect to a positive definite matrix $\qW \succ 0$ with $\qW = \qW^\top$ is denoted by $\norm{\qv}_{\qW} = \sqrt{\qv^\top \qW \qv}$. 
We denote the identity matrix of size $N \times N$ by $\qI_N$, and the zero matrix of suitable dimension by $\qZero$.
The Kronecker product of two matrices $\qA$ and $\qB$ is $\qA \otimes \qB$.
We recall that a function $\alpha : \qRealNumbers_{\geq 0} \rightarrow \qRealNumbers_{\geq 0}$ is of class $\Kfunc$ if it is continuous, strictly increasing, and satisfies $\alpha (0) = 0$. 
By $\Lfunc$, we refer to the class of functions $\theta : \qRealNumbers_{\geq 0} \rightarrow \qRealNumbers_{\geq 0}$ that are continuous, non-increasing, and satisfy $\lim_{s\rightarrow\infty} \theta(s) = 0$, and by $\KLfunc$ to the class of functions $\beta : \qRealNumbers_{\geq 0} \times \qRealNumbers_{\geq 0} \rightarrow \qRealNumbers_{\geq 0}$ with $\beta (\cdot, s) \in \Kfunc$ and $\beta (r, \cdot) \in \Lfunc$ for any fixed $s \in \qRealNumbers_{\geq 0}$ and $r \in \qRealNumbers_{\geq 0}$, respectively.
Last, we denote the space of all block-lower-triangular Toeplitz matrices of dimension $N$ with sub-matrices, $\forall n \in \qNaturalNumbers_{[1,N]}$, $\qTb_n \in \qRealNumbers^{L \times M}$, by $\mathcal{T}^{L,M}_{N}$, that is,
\begin{equation}
	\forall \qT \in \mathcal{T}_{N}^{L,M}, \quad \qT =  \begin{bmatrix}
		\qTb\idx{1} & \qvec{0} & \dots &   \qvec{0} \\
		\qTb\idx{2} & \qTb\idx{1} & \ddots & \qvec{0} \\
		\vdots & \ddots & \ddots & \vdots \\
		\qTb_N & \qTb_{N-1} & \dots & \qTb\idx{1}
	\end{bmatrix}\,.
\end{equation}

\section{Problem Statement}\label{sec:problem}
The central problem addressed in this paper is to find a control scheme that solves reference tracking problems in repetitive \ac{mimo} systems \emph{autonomously}, meaning that it does not require prior model information and that it does not require manual tuning for a specific system and reference.

Formally, consider a repetitive, discrete-time, \ac{mimo} system with input vector $\quv_j(n)\in\qRealNumbers^O$ and output vector $\qyv_j(n)\in\qRealNumbers^O$, where $j\in\qNaturalNumbersZero$ is the trial index, $n\in \qNaturalNumbers_{[1,N]}$ is the time sample index, and $N\in\qNaturalNumbersPos$ is the number of samples in a trial.
The samples of the input and output vectors are collected in the so-called input trajectory $\qu_j\in\qRealNumbers^{ON}$ and in the output trajectory $\qy_j\in\qRealNumbers^{ON}$, respectively, with $\forall j \in \qNaturalNumbersZero$,
\begin{align}
	\qu_j& \triangleq \begin{bmatrix}
		\quv^\top_j(1) & \dots & \quv^\top_j(N)
	\end{bmatrix}^\top\,, \label{eq:inputs} \\
	\qy_j& \triangleq \begin{bmatrix}
		\qyv^\top_j(1+p) & \dots & \qyv^\top_j(N+p)
	\end{bmatrix}^\top\,, \label{eq:outputs}
\end{align}
where $p\in\qNaturalNumbersZero$ is the system's relative degree that is assumed, without loss of generality, to be equal to one.

The system dynamics are assumed to be linear and time-invariant, 
such that the relation between the input and output trajectory can be described by the map
\begin{equation}\label{eq:ilc_dynamics}
	\foralljinN \qy_j = \qP\qu_j\,,
\end{equation}
where $\qP\in \mathcal{T}_N^{O,O}$ is the plant matrix, which has a block lower-triangular Toeplitz structure with the sub-matrices, $\forall n \in \qNaturalNumbers_{[1,N]}$, $\bar{\qP}_n \in \qRealNumbers^{O \times O}$, \ie
\begin{equation}\label{eq:P_matrix}
	\qP = \begin{bmatrix}
		\bar{\qP}\idx{1} & \qvec{0} & \dots &   \qvec{0} \\
		\bar{\qP}\idx{2} & \bar{\qP}\idx{1} & \ddots & \qvec{0} \\
		\vdots & \ddots & \ddots & \vdots \\
		\bar{\qP}_N & \bar{\qP}_{N-1} & \dots & \bar{\qP}\idx{1}
	\end{bmatrix}\,.
\end{equation}
Please note, the plant matrix $\qP$ is assumed to be \emph{unknown}. 

The objective is to find a learning scheme that autonomously determines a control input $\qu_j$ that steers the system output $\qy_j$ of an arbitrary \emph{unknown} \ac{mimo} system \eqref{eq:ilc_dynamics} to follow an arbitrary reference trajectory $\qr \in \qRealNumbers^{ON}$, without manual tuning for a specific system and or a specific reference.
To solve this problem, we set off from a \ac{mbilc} perspective and include an interactively learned system approximation to enable \ac{ddilc}. In particular, consider the following two sub-problems.


\subsection{Model Learning Problem}\label{sec:model_learning_problem}
The first sub-problem, referred to as the \emph{model learning problem}, consists of iteratively learning a model matrix $\qM_j \in \mathcal{T}_{N}^{O,O}$ to approximate the unknown plant matrix $\qP$. 
It is assumed that a sequence of input-output trajectory pairs $(\qu_j, \qy_j)$, stemming from the dynamics \eqref{eq:ilc_dynamics}, is available, and that the contained inputs are \emph{persistently exciting}.


\begin{Assumption}[Persistency of excitation]\label{as:excitation_condition}
	For any trial $j \in \qNaturalNumbersZero$, there exists a finite window of consecutive trials $\left[j, j + O -1\right]$, over which the first input samples $\quv_i(1)$, $\forall i \in \qNaturalNumbers_{[j,j+O-1]}$, collectively satisfy the excitation condition
	\begin{align}
		\foralljinN \mathrm{rank}\left(\begin{bmatrix}
			\quv_j(1) & \dots & \quv_{j+O-1}(1)
		\end{bmatrix}\right) = O \,.
	\end{align}
\end{Assumption}
Note, Assumption \ref{as:excitation_condition} refers to the linear independence of the initial input trajectory samples $\quv_j(1)$ over any finite window of $O$ trials. This ensures that the model learning problem is persistently well-posed after $O$ trials.


\subsection{Input Learning Problem}\label{sec:input_learning_problem}
The second problem, called the \emph{input learning problem}, consists of iteratively learning an input trajectory $\qu_j$ such that the output trajectory $\qy_j$ tracks a desired reference trajectory $\qr\in\qRealNumbers^{ON}$. 
Tracking performance is judged based on the error trajectory
\begin{equation}\label{eq:ilc_tracking_error}
	\foralljinN \qe_j  \triangleq  \qr - \qy_j\, ,
\end{equation}
and its Euclidean norm, which should ideally converge, in a monotonic fashion, to a value close to zero.




\section{Preliminaries: Iterative Learning Control}\label{sec:preliminaries}
In the following, we introduce relevant preliminaries on \acl{ilc} (\ac{ilc}) that build the foundation for the proposed methods in Section \ref{sec:methods}.

The core idea in \ac{ilc} is to iteratively learn a control input $\qu_j$ to make the system output $\qy$ follow a desired reference $\qr$, which corresponds to the input learning sub-problem \ref{sec:input_learning_problem}. To this end, the standard learning procedure consists of updating the control input based on the tracking error according to
\begin{equation} \label{eq:ilc_update}
	\foralljinN \qu_{j+1} = \qu_j + \qL_j \qe_j \, ,
\end{equation} 
in which $\qL_j$ is a -- possibly trial-varying -- learning gain matrix. To determine the learning gain matrix, different model-based or model-free \ac{ilc} designs can be employed \cite{Bristow.2006,Wang.2009}. In the following, we consider \ac{ilc} designs that rely on an (approximate) system model $\qM_j$. This class of \ac{mbilc} designs can be summarized by the generic \emph{design function} $\qDe : \mathcal{T}_{N}^{O,O} \rightarrow \qRealNumbers^{ON \times ON}$ with
\begin{equation} \label{eq:ilc_design_function}
	\foralljinN \qL_{j} = \qDe\left(\qM_j\right) \, .
\end{equation} 

\begin{Assumption}[Design function continuity]\label{as:design_function_continuity}
	It is assumed that all employed design functions $\qDe$ are continuous, that is, for every $\epsilon > 0$ there exists a $\delta > 0$ such that
	\begin{align}
		\norm{\qX - \qY} < \epsilon \quad \implies \quad \norm{\qDe\left(\qX\right) - \qDe\left(\qY\right)} < \delta \, .
	\end{align}
\end{Assumption}

Having set the core ingredients to solve an \ac{ilc} problem, we are in the position to define an \ac{ilc} system, its convergence properties, and typical design options.

\begin{Definition}[ILC system]
	The combination of dynamics \eqref{eq:ilc_dynamics} that allow for iterative experiments, error trajectory \eqref{eq:ilc_tracking_error}, update law \eqref{eq:ilc_update}, and design function \eqref{eq:ilc_design_function} is called an \textit{\ac{ilc} system}.
\end{Definition}

\begin{Definition}[Monotonic tracking convergence]\label{def:monotonic_tracking_convergence}
	An \ac{ilc} system is called \textit{exponentially monotonically tracking convergent} if and only if there exists $\alpha \in [0,1)$ such that
	\begin{align}
		\foralljinN \left\|\mathbf{e}_{j+1}\right\| \leq \alpha\left\|\mathbf{e}_j\right\| .
	\end{align}
\end{Definition}

\begin{Definition}[Gradient ILC]
	An \ac{ilc} system with design function
	\begin{align}\label{eq:design_gilc}
		\qDe_{\mathrm{G}}\left(\qM_j, \qW_j \right)=  \qM_j^{\top} \qW_j \, ,
	\end{align}
	in which $\qW_j \in \qRealNumbers^{ON \times ON}$ is a symmetric, positive definite matrix, is called \textit{gradient} \ac{ilc} (\acs{gilc}) \cite{Owens.2016,Chu.2025}.
\end{Definition}

\begin{Definition}[Norm-optimal ILC]
	An \ac{ilc} system with design function
	\begin{align}\label{eq:design_noilc}
		\qDe_{\mathrm{NO}}\left(\qM_j, \qW_j, \qS_j\right)=\left(\qM_j^{\top} \qW_j \qM_j+\qS_j\right)^{-1} \qM_j^{\top} \qW_j \, ,
	\end{align}
	in which $\qW_j \in \qRealNumbers^{ON \times ON}$ and $\qS_j \in \qRealNumbers^{ON \times ON}$ are symmetric, positive definite matrices, is called \textit{norm-optimal} \ac{ilc} (\acs{noilc}) \cite{Owens.2016,Chu.2025}.
\end{Definition}

\section{Proposed Method}\label{sec:methods}
In this section, two learning frameworks are presented to address the model and input learning problem in \ac{mimo} systems, respectively. 
The \acl{iml} (\acs{iml}) scheme considers the model learning problem.
The \acl{dilc} (\acs{dilc}) framework combines \ac{iml} with model-based \ac{ilc} to address the input learning problem.

The core idea is to transform the model learning problem such that established model-based \ac{ilc} methods can be utilized to iteratively approximate a model of the unknown dynamics. 
This relationship can be interpreted as a ``duality'' between \ac{ilc} and \ac{iml}. 
In the following, we adopt corresponding ideas from \cite{Meindl.2025, Liu.2016b}, and develop methods for the more complex and structurally different case of \ac{mimo} systems. 

\subsection{Iterative Model Learning}\label{sec:methods_iml}
To solve the model learning problem using established \ac{ilc} methods, consider the predicted output $\qyh_j(\qu) = \qM_j\qu$, given an input $\qu$, and note that the parameters $\qm_j \in \qRealNumbers^{O^2 N}$ contained in $\qM_j$ appear linearly in the prediction equation. 
Thus, noting that the prediction $\qyh_j(\qu)$ can be expressed in terms of the parameter vector $\qm_j$ using suitable transformations (to be defined later), the linear-in-the-parameters model learning problem can be interpreted as an \ac{ilc} problem \cite{Meindl.2025}. 
To transform the model approximation $\qM_j$ and the input vector $\qu$ accordingly, two lifting operators are introduced.

To this end, for any given input vector $\qu$,
let $\qLu : \qRealNumbers^{ON} \rightarrow {\mathcal{T}}_{N}^{O,O^2}$ denote the input lifting operator, defined by
\begin{equation}\label{eq:lifting_input}
	\forall \qu \in \qRealNumbers^{ON}, \quad  \qLu(\qu) = \begin{bmatrix}
		\qUb\idx{1} & \qvec{0} & \dots &   \qvec{0} \\
		\qUb\idx{2} & \qUb\idx{1} & \ddots & \qvec{0} \\
		\vdots & \ddots & \ddots & \vdots \\
		\qUb_N & \qUb_{N-1} & \dots & \qUb\idx{1}
	\end{bmatrix}\,,
\end{equation}
with, $ \forall n \in \qNaturalNumbers_{[1,N]}$, 
\begin{align}
		\qUb_n &= \qI_{O} \otimes \quv^\top(n) \equiv \begin{bmatrix}
					\quv^\top(n)			& \qZero & \dots  & \qZero \\
					\qZero	& \quv^\top(n)		 & \ddots & \qZero \\
					\vdots          	& \ddots		     & \ddots & \vdots  \\
					\qZero  & \qZero & \dots  & \quv^\top(n)
				\end{bmatrix} \, .
\end{align}

Similarly, for any given model or plant matrix $\qM_j, \qP  \in \mathcal{T}_{N}^{O,O}$, let $\qLm : \mathcal{T}_{N}^{O,O} \rightarrow \qRealNumbers^{O^2 N}$ denote the model lifting operator, defined by
\begin{equation}\label{eq:lifting_model}
	\forall \qP \in \mathcal{T}_{N}^{O,O}, \quad \qLm (\qP) = \begin{bmatrix}
		\mathrm{vec}(\bar{\qP}\idx{1}) \\
		\mathrm{vec}(\bar{\qP}\idx{2}) \\
		\vdots \\
		\mathrm{vec}(\bar{\qP}_N) 
	\end{bmatrix} \, ,
\end{equation}
that transforms the model matrix $\qM_j$, or plant matrix $\qP$, into the parameter vectors $\qm_j, \qp \in\qRealNumbers^{O^2 N}$, respectively.

\begin{Remark}
	This paper considers square \ac{mimo} linear time-invariant systems, resulting in a model vector of size $O^2 N$. However, the presented theory extends to topologies of $\qP$ other than \eqref{eq:P_matrix}, \eg to reflect linear time-varying dynamics or to embed other prior knowledge about the system's structure. This can be done by designing suitable lifting operations and adjusting the model vector size accordingly. 
\end{Remark}

Note that the respective inverses of $\qLu$ and $\qLm$ exist.

\begin{Lemma}\label{lem:lifting_operations}
	Given a block-lower-triangular Toeplitz matrix $\qA \in \mathcal{T}_{N}^{O,O}$ and a vector $\qx \in \qRealNumbers^{ON}$, then
	\begin{equation}
		\qA\qx = \qLu(\qx)\qLm(\qA)\,.
	\end{equation}
\end{Lemma}
\begin{IEEEproof}
	Let 
	\begin{equation}
		\qw =  \begin{bmatrix}
			\qw_1 \\ \qw_2 \\ \vdots \\ \qw_N
		\end{bmatrix} = \begin{bmatrix}
			\qA\idx{1} & \qvec{0} & \dots &   \qvec{0} \\
			\qA\idx{2} & \qA\idx{1} & \ddots & \vdots \\
			\vdots & \ddots & \ddots & \vdots \\
			\qA_N & \qA_{N-1} & \dots & \qA\idx{1}
		\end{bmatrix} \begin{bmatrix}
			\qx_1 \\ \qx_2 \\ \vdots \\ \qx_N
		\end{bmatrix} = \qA\qx,
	\end{equation}
	with vectors, $\forall n \in \qNaturalNumbers_{[1,N]}$, $\qw_n, \qx_n \in \qRealNumbers^{O}$, and sub-matrices $\qA_n \in \qRealNumbers^{O\times O}$, respectively, then the $k$-th component of the vector $\qw_n$ is given by
	\begin{equation}
		\begin{aligned}
		[ \qw_{n} ]_k &= \sum_{j=1}^{n}  [\qA_j \qx_{n-j+1} ]_{k} 
		= \sum_{j=1}^{n}   \left[ \left( \qI_{O} \otimes \qx_{n-j+1}^\top \right) \mathrm{vec}(\qA_j)  \right]_{k},
		\end{aligned}
	\end{equation}
	and, hence, 
	\begin{equation}
		\qA\qx = \qw = \qLu(\qx)\qLm(\qA)\,.
	\end{equation}
\end{IEEEproof}

Using the lifting operations, we can introduce the following variables. 
Let $\qm_j=\qLm(\qM_j)$ denote the model vector, let $\qp=\qLm(\qP)$ denote the plant vector, and let $\qU=\qLu(\qu)$ denote the input matrix. Based on Lemma \ref{lem:lifting_operations}, the predicted output trajectory $\qyh$ for some input trajectory $\qu$ is given by
\begin{equation}\label{eq:iml_prediction}
	\foralljinN \qyh_j(\qu) \triangleq \qM_j\qu \equiv \qU \qm_j\,.
\end{equation}
The prediction error $\qeh_j$ based on the model estimate in trial $j$ is defined as the difference between some output trajectory $\qy$ and the predicted output trajectory $\qyh_j(\qu)$ with corresponding input $\qu$, \ie
\begin{equation}\label{eq:iml_pred_error}
	\foralljinN  \qeh_j(\qy, \qu) \triangleq  \qy-\qU \qm_j\,,
\end{equation}
and the model vector is iteratively updated by
\begin{equation}\label{eq:iml_update}
	\foralljinN \qm_{j+1} = \qm_j + \qLh_j\qeh_j(\qy_j, \qu_j)\,,
\end{equation}
where $\qLh_j \in \qRealNumbers^{O^2 N \times ON}$ is the \ac{iml}'s trial-varying learning-gain matrix that is determined by the design function $\qDeh : \mathcal{T}_{N}^{O,O^2} \rightarrow \qRealNumbers^{O^2 N \times ON}$ with
\begin{equation}\label{eq:iml_design_function}
	\foralljinN \qLh_j = \qDeh(\qU_j)\,.
\end{equation}
Equipped with these characterizations of the model learning procedure, an \ac{iml} system can be formally defined.

\begin{Definition}[IML system]
	The combination of prediction dynamics \eqref{eq:iml_prediction}, prediction error trajectory \eqref{eq:iml_pred_error}, model update law \eqref{eq:iml_update}, and design function \eqref{eq:iml_design_function} is called an \textit{\ac{iml} system}.
\end{Definition}

Based on the equivalence established in \eqref{eq:iml_prediction}, an \ac{iml} system can be interpreted as an \ac{ilc} system with trial-varying reference and trial-varying, but \emph{unknown} dynamics \cite[Theorem 1]{Meindl.2025}. 
Specifically, the duality consists in the correspondence between, for instance, the \ac{ilc} output dynamics \eqref{eq:ilc_dynamics} and the \ac{iml} prediction dynamics \eqref{eq:iml_prediction}, or the \ac{ilc} tracking error \eqref{eq:ilc_tracking_error} and the \ac{iml} prediction error \eqref{eq:iml_pred_error}. 

Exploiting the duality between \ac{ilc} and \ac{iml}, established model-based \ac{ilc} methods can be employed for iterative learning of system models. 
In particular, established \ac{ilc} design functions can be used to (autonomously) construct the learning gain matrix $\qLh$. 
In this paper, we consider two of the arguably most common schemes for model-based design of learning gain matrices, \ac{gilc} and \ac{noilc} \cite{Owens.2016,Chu.2025}, and exploit them for \ac{iml}. The resulting design functions are
\begin{align}\label{eq:design_giml}
	\qDeh_{\mathrm{G}}\left(\qU_j, \qW_j \right)=  \qU_j^{\top} \qW_j \, ,
\end{align}
for \ac{giml}, and
\begin{align}\label{eq:design_noiml}
	\qDeh_{\mathrm{NO}}\left(\qU_j, \qW_j, \qS_j\right)=\left(\qU_j^{\top} \qW_j \qU_j+\qS_j\right)^{-1} \qU_j^{\top} \qW_j \, ,
\end{align}
for \ac{noiml}.



\subsection{Dual Iterative Learning Control}\label{sec:methods_dilc}
In this section, the input learning problem is considered. It consists of iteratively learning an input trajectory $\qu_j$ such that the output trajectory $\qy_j$ tracks a desired reference trajectory $\qr\in\qRealNumbers^{ON}$, without knowledge about the plant dynamics. 
To solve this task, we present \ac{mimo} \acl{dilc} (\ac{dilc}), which uses the approximated model obtained from \ac{iml} within each \ac{ilc} trial. 
The framework enables simultaneous iterative learning of both the control input and the system model. 
The duality of the \ac{iml} and \ac{ilc} systems enables the use of various established model-based \ac{ilc} schemes in either the model learning or the control learning part. 

Specifically, each iteration of \ac{dilc} consists of three steps, depicted in Figure \ref{fig:architecture}. First, an experimental trial with control input $\qu_j$ is carried out, and the corresponding system output $\qy_j$ is measured. Second, the trajectory pair $(\qu_j, \qy_j)$ is used for \textit{model learning} by \ac{iml}, which consists of applying the corresponding lifting operations, designing the learning gain, and updating the model vector based on \eqref{eq:iml_update}. Third, \textit{control learning} is performed by employing the updated model in model-based \ac{ilc}, again performing lifting, learning gain computation, and executing the update law \eqref{eq:ilc_update} to determine the next control input $\qu_{j+1}$. The resulting procedure is summarized in Algorithm \ref{algo:dilc}.


\begin{algorithm}
	\caption{Dual Iterative Learning Control (\ac{dilc})}
	\label{algo:dilc}
	\begin{algorithmic}[1]
		\State \textbf{Input:} Reference trajectory $\qr$, number of trials $J$.
		\State \textbf{Initialize:} Control input $\qu_0$, model vector $\qm_0$, and design functions $\qDe(\cdot)$, $\qDeh(\cdot)$.
		\For{$j=0$ \textbf{to} $J-1$}
		\State Apply input $\qu_j$ to the system and measure output $\qy_j$.
		\Statex \hspace{5mm} \textit{// Model Learning} \Comment{See Section \ref{sec:methods_iml}}
		\State $\qU_j \leftarrow \qLu(\qu_j)$ \Comment{by \eqref{eq:lifting_input}}
		\State $\qLh_j \leftarrow \qDeh(\qU_j)$ \Comment{by \eqref{eq:iml_design_function}}
		\State $\qm_{j+1} \leftarrow \qm_j + \qLh_j \qeh_j (\qy_j, \qu_j)$ \Comment{by \eqref{eq:iml_update}}
		\Statex \hspace{5mm} \textit{// Control Learning} \Comment{See Section \ref{sec:preliminaries}}
		\State $\qM_{j+1} \leftarrow \qLm^{-1}(\qm_{j+1})$ \Comment{by \eqref{eq:lifting_model}}
		\State $\qL_j \leftarrow \qDe(\qM_{j+1})$ \Comment{by \eqref{eq:ilc_design_function}}
		\State $\qu_{j+1} \leftarrow \qu_j + \qL_j \qe_j$ \Comment{by \eqref{eq:ilc_update}}
		\EndFor
		\State \textbf{return} Learned control input $\qu_J$.
	\end{algorithmic}
\end{algorithm}

\subsection{Convergence Analysis}\label{sec:methods_convergence}
In this section, the theoretical properties of the presented \ac{iml} and \ac{dilc} methods are analyzed. For \ac{iml}, we derive convergence conditions of the model approximation error and the model prediction error, based on the duality to \ac{ilc} algorithms. For \ac{dilc}, conditions for tracking error convergence without model knowledge are provided.

In the context of \ac{iml}, the core quantity of interest is the model approximation error between the model vector $\qm_j$ and the true plant parameter vector $\qp = \qLm (\qP)$, that is $\qem_j = \qp - \qm_j$. The progression of this error quantity along the trial dimension can be characterized by the notion of \textit{monotonic model convergence}.

While the model error can be shown to converge strictly monotonically in \ac{siso} dynamics, the model learning problem is overparameterized in repetitive \ac{mimo} systems. Specifically, $O^2 N$ model parameters need to be inferred, but each measurement trajectory provides only $ON$ constraints. Given Assumption \ref{as:excitation_condition}, the parameter inference problem becomes well-posed only if $O$ consecutive trials are considered jointly. Thus, to account for arbitrary error decay rates, \ie that do not necessarily converge \textit{exponentially}, the following definition for the model convergence is based on general $\KLfunc$-functions.

\begin{Definition}[Monotonic model convergence]\label{def:monotonic_model_convergence}
	An \ac{iml} system is called \textit{monotonically model convergent} if and only if there exists a function $\beta \in \KLfunc$ such that, $\forall j \in \qNaturalNumbersPos$, 
	\begin{equation}
		\norm{\qem_j } \leq \beta \left(\norm{\qem_0},j \right) \leq \norm{\qem_{j-1} } \, .
	\end{equation}
\end{Definition}


\begin{Theorem}[Monotonic model convergence]\label{thm:model_convergence}
	Given Assumption \ref{as:excitation_condition}, a \ac{mimo} \ac{iml} system is \textit{monotonically model convergent} after $O$ trials {if}
	\begin{equation}\label{eq:thm1_condition}
		\foralljinN \norm{{\qI} - \qLh_j \qU_j} \leq 1 \, ,
	\end{equation}
    and all learning gain matrices $\qLh_j$ have full column rank.
    \begin{IEEEproof}
        The proof is shown in Appendix \ref{app:proof_thm_model_conv}.
    \end{IEEEproof}
\end{Theorem}



A second quantity of interest, essential for the \ac{iml} learning update, is the prediction error $\qeh_j(\qy,\qu)$. In the following, we judge its dynamics by the notion of \textit{monotonic prediction convergence}.

\begin{Definition}[Monotonic prediction convergence]
	An \ac{iml} system is said to be \textit{exponentially monotonically prediction convergent} if and only if there exists a $\gamma \in [0,1)$ such that
	\begin{equation}
		\foralljinN \norm{\qeh_{j+1}(\qy_{j}, \qu_{j}) } \leq \gamma \norm{\qeh_j(\qy_{j}, \qu_{j})} \, .
	\end{equation}
\end{Definition}

\begin{Theorem}[Monotonic prediction convergence]\label{thm:prediction_convergence}
	A \ac{mimo} \ac{iml} system is \textit{exponentially monotonically prediction convergent} if there exists a fixed $\gamma \in [0,1)$ such that
	\begin{equation}\label{eq:thm2_condition}
		\foralljinN \norm{{\qI} - \qU_j \qLh_j} \leq \gamma \, .
	\end{equation}
	\begin{IEEEproof}
		Combining \eqref{eq:iml_prediction}-\eqref{eq:iml_update} yields
		\begin{equation}
			\foralljinN \qeh_{j+1}(\qy_{j}, \qu_{j}) = \left( \qI -  \qU_j \qLh_j \right) \qeh_{j}(\qy_{j}, \qu_{j}) \, .
		\end{equation}
		The norm of the right-hand side can be upper-bounded using the inequality, $\foralljinN $
		\begin{equation}
			\norm{ \left(\qI -  \qU_j \qLh_j \right) \qeh_{j}(\qy_{j}, \qu_{j})} \leq \norm{ \qI -  \qU_j \qLh_j } \norm{\qeh_{j}(\qy_{j}, \qu_{j})}   \, .
		\end{equation}
		For monotonic prediction convergence, the term $\qI -  \qU_j \qLh_j$ is required to be a contraction with a fixed upper bound $\gamma \in [0,1)$, \ie
		\begin{equation}
			\foralljinN \norm{ \qI -  \qU_j \qLh_j} \leq \gamma \, .
		\end{equation}
	\end{IEEEproof}
\end{Theorem}

Comparing Theorem \ref{thm:model_convergence} and Theorem \ref{thm:prediction_convergence}, it is clear that both relate to the convergence of model quality, but that small prediction errors could occur even if the model error $\qp - \qm_j$ is significant. 
This discrepancy becomes visible only in the \ac{mimo} case and relates to the overparametrization of the model representation \eqref{eq:P_matrix}.
In this context, \textit{model convergence} can be interpreted as a stronger convergence notion than \textit{prediction convergence}.

Considering \ac{dilc}, the main quantity of interest is the tracking error \eqref{eq:ilc_tracking_error}, which relates to the control learning problem. Its progression can be characterised by the concept of \textit{monotonic tracking convergence}, according to Definition \ref{def:monotonic_tracking_convergence}.

\begin{Theorem}[Monotonic tracking convergence]\label{thm:tracking_convergence}
	A \ac{mimo} \ac{dilc} system with model learning design function $\qDeh$ and control learning design	function $\qDe$ is monotonically model convergent, and there exists a threshold trial $J \in \qNaturalNumbersZero$ from which onward the control learning is exponentially monotonically tracking convergent, \ie there exists $\alpha \in [0,1)$ such that
	\begin{equation}
		\forall j \in \qNaturalNumbers_{\geq J}, \quad \norm{\qe_{j+1} } \leq \alpha \norm{\qe_j} \, ,
	\end{equation}
	if the design function $\qDeh$ is continuous and leads to monotonic model convergence in a \ac{mimo} \ac{iml} system, and the design function $\qDe$ is continuous and leads to exponentially monotonic tracking convergence in a \ac{mimo} \ac{ilc} system.
	\begin{IEEEproof}
        The following proof is a direct extension of the proof of Theorem 5 in \cite{Meindl.2025} for the case of \ac{mimo} dynamics. 
        Given that the design function $\qDe$ leads to exponentially monotonic tracking convergence in an \ac{ilc} system with a known system matrix $\qP$, and that the design functions are continuous, there exists a neighbourhood around $\qP$ for which model approximations lead to monotonic tracking convergence.
		Provided that the design function $\qDeh$ leads to monotonic model convergence in an \ac{iml} system, the model approximation $\qM_j$ enters this neighborhood of $\qP$ within a finite number of trials $J \in \qNaturalNumbersZero$, implying exponentially monotonic tracking convergence from the threshold trial $J$ onwards.
	\end{IEEEproof}
\end{Theorem}

To ensure that the conditions \eqref{eq:thm1_condition} and \eqref{eq:thm2_condition} are met, suitable \ac{iml} design functions are required. 
It can be shown that condition \eqref{eq:thm1_condition} for monotonic model convergence is always satisfied in \ac{noiml} and \ac{noilc} due to the symmetric structure of $\qLh_j\qU_j$ and $\qL_j\qU_j$, and the weighting matrix $\qS_j$ being positive definite. 
For prediction convergence in \ac{noiml}, \ie for condition \eqref{eq:thm2_condition} to hold, the weighting matrices $\qW_j$ and $\qS_j$ need to be tuned. 
Similarly, in \ac{giml}, the conditions \eqref{eq:thm1_condition} and \eqref{eq:thm2_condition} are satisfied only subject to a sufficiently small $\qW_j$, which induces a trade-off between convergence speed and learning stability in the trial dimension. 
This raises the question of systematic rules for tuning of $\qW_j$ and $\qS_j$.

\subsection{Self-Parametrization}\label{sec:methods_self_parametrization}
In this section, we consider strategies for self-parametrization of the \ac{dilc} schemes to enable reliable autonomous learning of different reference tracking tasks across various systems without manual tuning. 
Following the previous derivations, we focus on gradient-based and norm-optimal design functions.

To this end, two problems need to be addressed by suitable tuning rules for the weighting parameters $\qW_j$ and $\qS_j$. First, the choice of $\qW_j$ and $\qS_j$ critically influences the error convergence properties in \ac{ilc} and \ac{iml} schemes, potentially leading to diverging error terms and system failure. Second, without model information, the scale between different inputs and outputs can vary significantly, leading to either very slow or excessively high learning speeds in different input or output dimensions.

To account for the latter, we propose considering the relationships between each input and each output separately, treating the \ac{mimo} system with $O$ inputs and $O$ outputs as a superposition of $O^2$ \ac{siso} systems, each described by a lower-triangular Toeplitz matrix of size $N \times N$.



\begin{Lemma}[Superposition representation]\label{lem:superposition_representation}
	An equivalent representation of the input-output relationship \eqref{eq:ilc_dynamics} is given by grouping the samples by input and output dimension, instead of grouping them by sampling time. The equivalent notation is denoted by the superscript $\tilde{\square}$, and reads $\tilde{\qy}_j = \tilde{\qP} \tilde{\qu}_j$ with
	\begin{align}
		\tilde{\qu}_j& \triangleq \begin{bmatrix}
			\bar{u}_{j,1}(1) \\ \bar{u}_{j,1}(2) \\ \vdots \\ \bar{u}_{j,1}(N) \\ \bar{u}_{j,2}(1) \\ \vdots\\ \bar{u}_{j,O}(N-1) \\ \bar{u}_{j,O}(N)
		\end{bmatrix}\,,  \quad \tilde{\qy}_j \triangleq \begin{bmatrix}
			\bar{y}_{j,1}(2) \\ \bar{y}_{j,1}(3) \\ \vdots \\ \bar{y}_{j,1}(N+1) \\ \bar{y}_{j,2}(2) \\ \vdots\\ \bar{y}_{j,O}(N) \\ \bar{y}_{j,O}(N+1)
		\end{bmatrix}\,.  
	\end{align}
	The input-output map $\tilde{\qP} \in \qRealNumbers^{ON \times ON}$ consists of $O^2$ sub-matrices with lower-triangular Toeplitz structure $\tilde{\qP}_{k,i} \in \mathcal{T}_{N}^{1,1}$, $i \in \qNaturalNumbers_{[1,O]}$, $k \in \qNaturalNumbers_{[1,O]}$, and reads
	\begin{align}
		\tilde{\qP} = \begin{bmatrix}
		\tilde{\qP}\idx{1,1} & \tilde{\qP}\idx{1,2} & \dots &   \tilde{\qP}_{1,O} \\
		\tilde{\qP}\idx{2,1} & \tilde{\qP}\idx{2,2} & \ddots & \vdots \\
		\vdots & \ddots & \ddots & \vdots \\
		\tilde{\qP}_{O,1} & \tilde{\qP}_{O,2} & \dots & \tilde{\qP}_{O,O}
	\end{bmatrix}\,,
	\end{align}
    where each input-output map $\tilde{\qP}_{k,i}$ represents \ac{siso} subsystem dynamics from input $i$ to output $k$, respectively.
\end{Lemma}

\begin{Remark}\label{rem:iml_superposition}
	Analogously to Lemma \ref{lem:superposition_representation}, the prediction dynamics \eqref{eq:iml_prediction}, employed in \ac{iml}, can be expressed as $\hat{\qyt}_j = \tilde{\qU} \tilde{\qm}_j$ with
	\begin{align}
		\hat{\qyt}_j \triangleq \begin{bmatrix}
			\hat{\tilde{y}}_{j,1}(2) \\ \hat{\tilde{y}}_{j,1}(3) \\ \vdots \\ \hat{\tilde{y}}_{j,1}(N+1) \\ \hat{\tilde{y}}_{j,2}(2) \\ \vdots\\ \hat{\tilde{y}}_{j,O}(N) \\ \hat{\tilde{y}}_{j,O}(N+1)
		\end{bmatrix}\,, \quad \tilde{\qm}_j & \triangleq \begin{bmatrix}
			\tilde{m}_{j,1}(1) \\ \tilde{m}_{j,1}(2) \\ \vdots \\ \tilde{m}_{j,1}(N) \\ \tilde{m}_{j,2}(1) \\ \vdots\\ \tilde{m}_{j,O^2}(N-1) \\ \tilde{m}_{j,O^2}(N)
		\end{bmatrix} \,,
    \end{align}
    and
    \begin{align}
		\tilde{\qU} &= \begin{bmatrix}
			\tilde{\qU}\idx{1,1} & \tilde{\qU}\idx{1,2} & \dots &   \tilde{\qU}_{1,O^2} \\
			\tilde{\qU}\idx{2,1} & \tilde{\qU}\idx{2,2} & \ddots & \vdots \\
			\vdots & \ddots & \ddots & \vdots \\
			\tilde{\qU}_{O,1} & \tilde{\qU}_{O,2} & \dots & \tilde{\qU}_{O,O^2}
		\end{bmatrix}\,,
	\end{align}
    where each submatrix $\tilde{\qU}_{k,i}$ represents the relationship between model parameter $i$ and predicted output dimension $k$, respectively.
\end{Remark}

Exploiting Lemma \ref{lem:superposition_representation} and Remark \ref{rem:iml_superposition}, we propose to parametrize the weighting matrix $\qW \in \qRealNumbers^{ON \times ON}$ in \ac{gilc} and \ac{giml} by
\begin{align}
	\foralljinN \qW_j &= \mathrm{diag} \left( \begin{bmatrix}
		\tilde{\qw}_{1} \\ \vdots \\ \tilde{\qw}_{N}
	\end{bmatrix} \right) , 
\end{align}
with the $i$-th entry of each $\tilde{\qw} \in \qRealNumbers^{O}$ being
\begin{equation}
    \begin{aligned}
	\tilde{w}_{i}^{\mathrm{ILC}} &= \frac{1}{\norm{\left[\tilde{\qM}_{i,1} \, \hdots \, \tilde{\qM}_{i,O} \right]}^2 } \, , \\
	\tilde{w}_{i}^{\mathrm{IML}} &= \frac{1}{\norm{\left[\tilde{\qU}_{i,1} \, \hdots \, \tilde{\qU}_{i,O^2} \right]}^2 } \, , \label{eq:selfpara_w_iml_eu}
    \end{aligned}
\end{equation}
for \ac{ilc} and \ac{iml}, respectively. 


For self-reliant parametrization of \ac{noilc} and \ac{noiml}, we propose to construct the weighting matrices $\qQ$, $\qS^{\mathrm{ILC}}$, and $\qS^{\mathrm{IML}}$ according to, $\foralljinN$
\begin{equation}
    \begin{aligned}
	\qQ_j = \mathrm{diag} & \left( \begin{bmatrix}
		\tilde{\qq}_{1} \\ \vdots \\ \tilde{\qq}_{N}
	\end{bmatrix} \right) ,  \\
	\qS_j^{\mathrm{ILC}} = \mathrm{diag} \left( \begin{bmatrix}
	\tilde{\qs}_{1}^{\mathrm{ILC}} \\ \vdots \\ \tilde{\qs}_{N}^{\mathrm{ILC}}
	\end{bmatrix} \right) , &\: 	\qS_j^{\mathrm{IML}} = \mathrm{diag} \left( \begin{bmatrix}
	\tilde{\qs}_{1}^{\mathrm{IML}} \\ \vdots \\ \tilde{\qs}_{N}^{\mathrm{IML}}
\end{bmatrix} \right) ,
    \end{aligned}
\end{equation}
with the $i$-th entry of each $\tilde{\qq} \in \qRealNumbers^{O}$, $\tilde{\qs}^{\mathrm{ILC}} \in \qRealNumbers^{O}$, and $\tilde{\qs}^{\mathrm{IML}} \in \qRealNumbers^{O^2}$ being
\begin{equation}
    \begin{aligned}
	\tilde{q}_{i}^{\mathrm{ILC}} &= \frac{1}{\norm{\left[\tilde{\qM}_{i,1} \, \hdots \, \tilde{\qM}_{i,O} \right]} } , \: 	\tilde{q}_{i}^{\mathrm{IML}} = \frac{1}{\norm{\left[\tilde{\qU}_{i,1} \, \hdots \, \tilde{\qU}_{i,O^2} \right]} } ,\\
    \tilde{s}_{i}^{\mathrm{ILC}} &= \norm{ \left[\tilde{\qM}_{1,i} \, \hdots \, \tilde{\qM}_{O,i} \right]} , \:     \tilde{s}_{i}^{\mathrm{IML}} = \norm{ \left[\tilde{\qU}_{1,i} \, \hdots \, \tilde{\qU}_{O,i} \right]}  . \label{eq:selfpara_s_iml_eu}
    \end{aligned}
\end{equation}

Intuitively, the self-parametrizations $\qW$ and $\qQ$ weight each tracking or prediction error with the inverse output sensitivity of the model, which addresses both varying magnitudes of the error terms as well as a stabilized learning behavior with regard to conditions \eqref{eq:thm1_condition} and \eqref{eq:thm2_condition}.
Further, in \ac{noilc} and \ac{noiml}, the weighting matrices $\qS^{\mathrm{ILC}}$ and $\qS^{\mathrm{IML}}$ penalize large step sizes in $\qu$ and $\qm$, respectively. 
Through the self-parametrization with the input sensitivity of the model, the relative step size in each input (\ac{noilc}) and parameter dimension (\ac{noiml}) is balanced. 

The proposed self-parametrization schemes are heuristics to trade off convergence speed and learning stability. Given the presented schemes, reliable learning capabilities of \ac{dilc} can be shown in a wide range of scenarios. Specifically, the model learning and input learning problems (see Section \ref{sec:problem}) are solved autonomously for various reference tracking tasks and across multiple systems in the sense that \textit{no model information} \textit{nor manual tuning} is required. These autonomous learning capabilities are illustrated in the following results section.

\section{Results}\label{sec:results}
In this section, the presented \ac{dilc} algorithm is tested in simulations and real-world experiments. 
We have verified the provided theorems in extensive simulation studies with linear systems. 
Notably, the proposed method---rapidly and autonomously---solves reference tracking tasks in nonlinear \ac{mimo} real-world systems, despite the underlying assumption of deterministic linear time-invariant dynamics. 
An interpretation is that the \ac{dilc} scheme learns an approximation to the linear time-varying system representation along the reference trajectory. 
In the following, we demonstrate the learning performance of \ac{dilc} in various real-world \ac{mimo} systems with different reference trajectories.
In particular, we consider reference tracking tasks in three systems,
\begin{enumerate}
    \item a 6-axis industrial robot simulation, using an UR10e \href{https://mujoco.readthedocs.io/en/stable/models.html}{MuJoCo} model,
    \item a real-world planar two-link robot, and
    \item a real-world balancing robot with two inputs and outputs, respectively.
\end{enumerate}

In each system, three different reference tracking tasks, each characterized by a known reference trajectory, are to be solved. 
In all simulations and experiments, the system model is unknown, and no manual parameter tuning is done. 
Instead, the self-parametrization strategies presented in Section \ref{sec:methods_self_parametrization} are employed. 
The learning performance is judged by the normalized tracking error norm, which is the current trial's tracking error norm, normalized by the first trial's tracking error norm, \ie $\norm{\qe_j} / \norm{\qe_0}$.

To perform \ac{dilc}, design functions need to be selected for both the \ac{iml} and the \ac{ilc} subsystem. 
To show that learning can be accomplished with different established model-based design functions, we tested various \ac{ilc} and \ac{iml} design function choices and combinations.
While the results of the experimental case studies are presented for a selection of different design functions to maintain readability, we emphasize that other design function choices also lead to rapid convergence.

In the following, we refer to a \ac{dilc} scheme with a certain design function choice by first naming the \ac{iml} design function and second naming the \ac{ilc} design function, \ie \ac{dilc} with \ac{giml} and \ac{noilc} is referred to as \acs{gnodilc}.


\subsection{6-Axis Robot Simulations}\label{sec:results_ur10e}

\begin{figure}[t]
	\centering
	\includegraphics[width=0.49\linewidth]{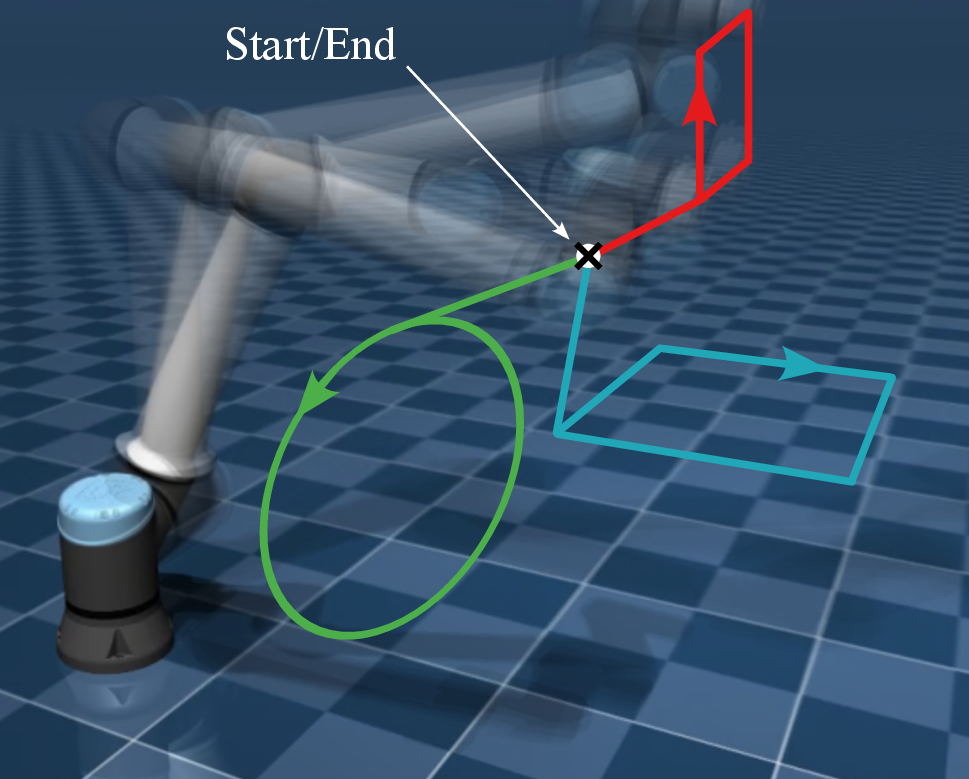}
	\includegraphics[width=0.49\linewidth]{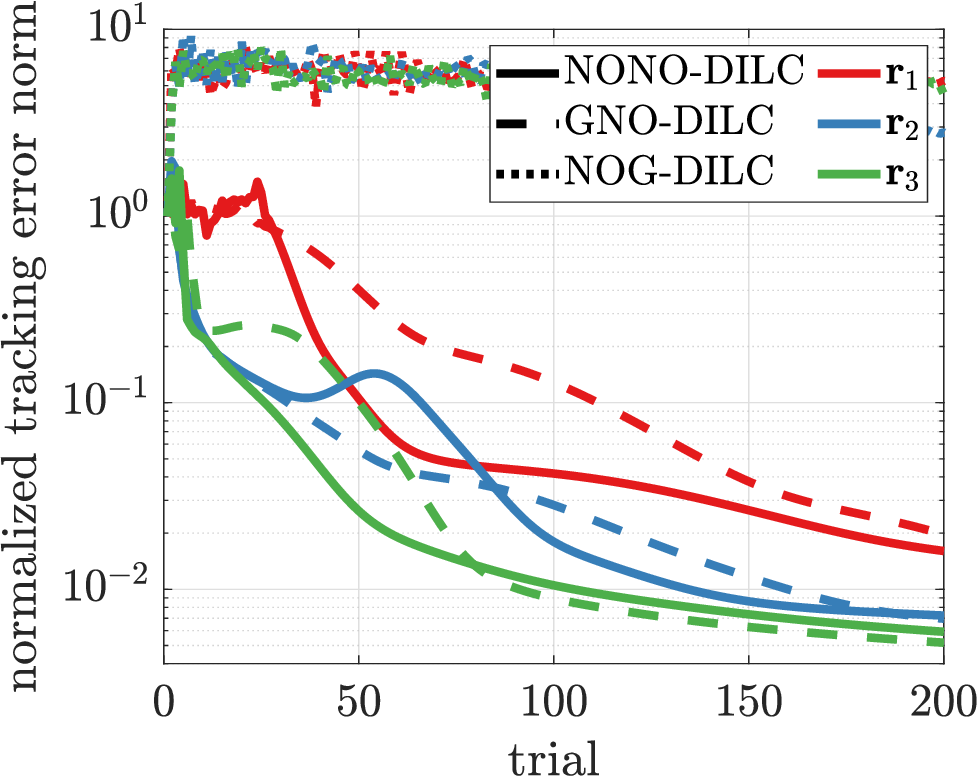}\hfill
	\vspace{2mm}
	\includegraphics[width=0.32\linewidth]{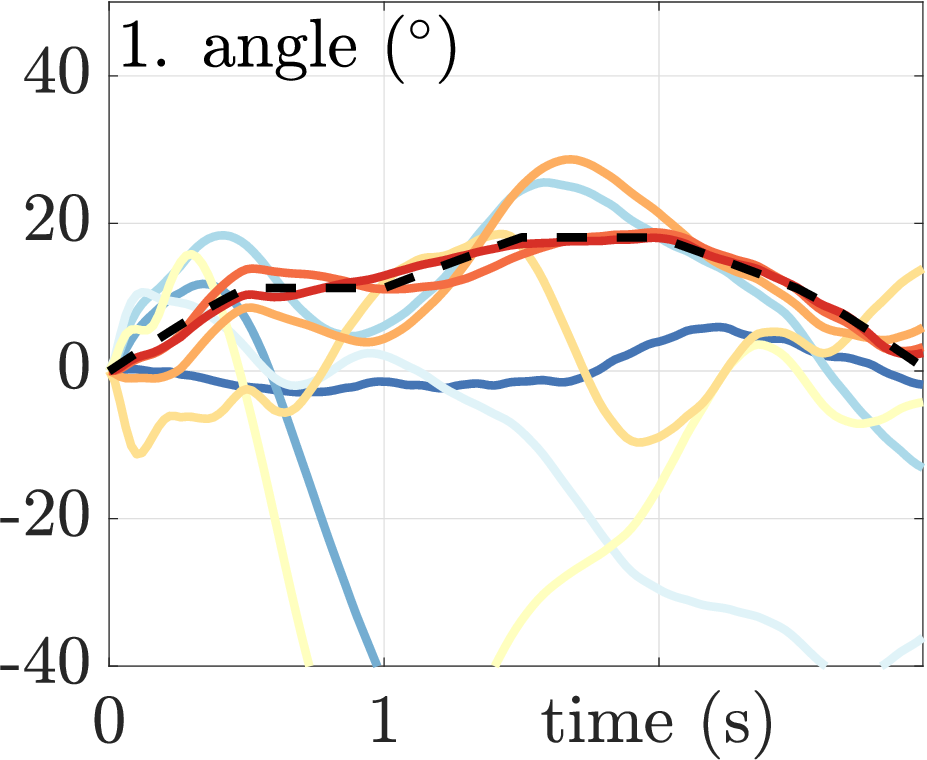}
	\includegraphics[width=0.32\linewidth]{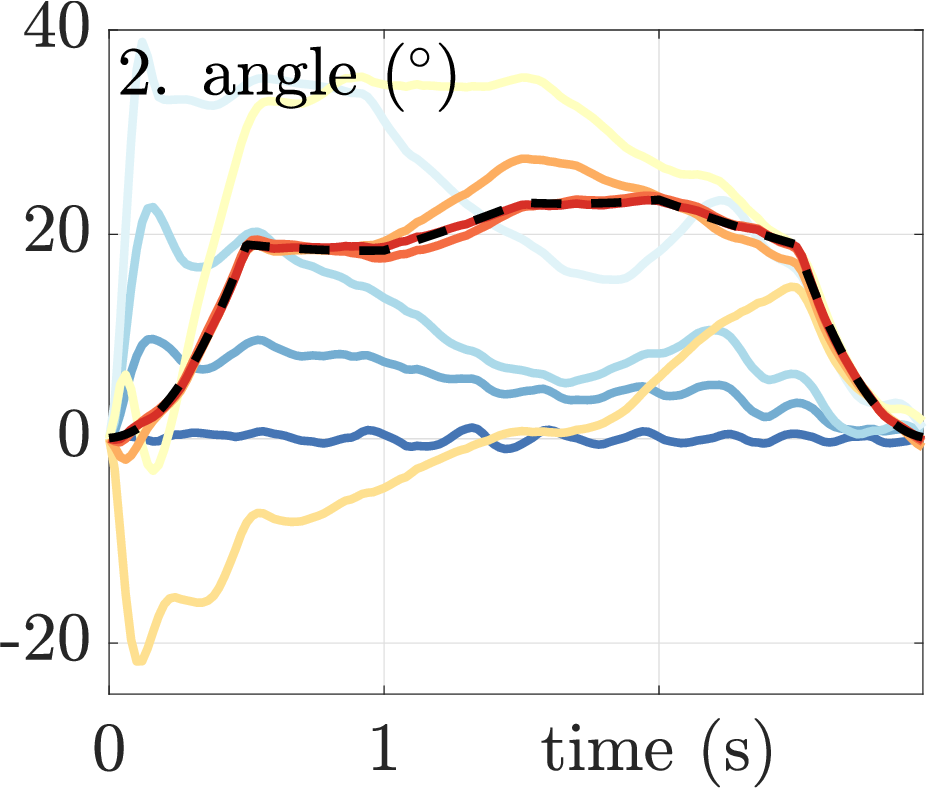}
	\includegraphics[width=0.32\linewidth]{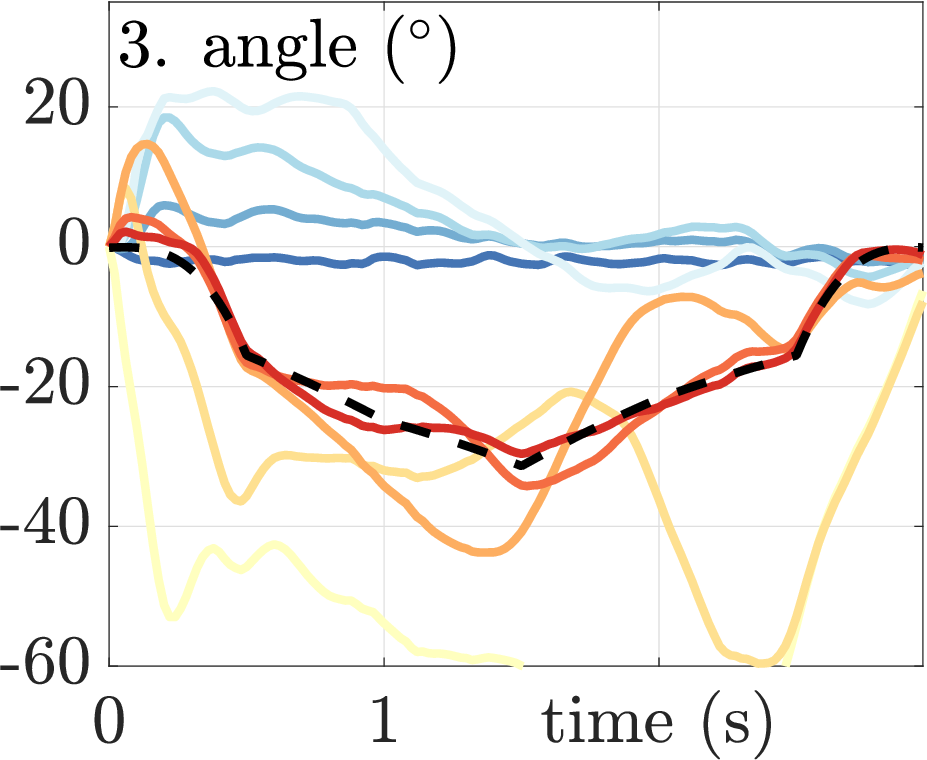}
	\includegraphics[width=0.32\linewidth]{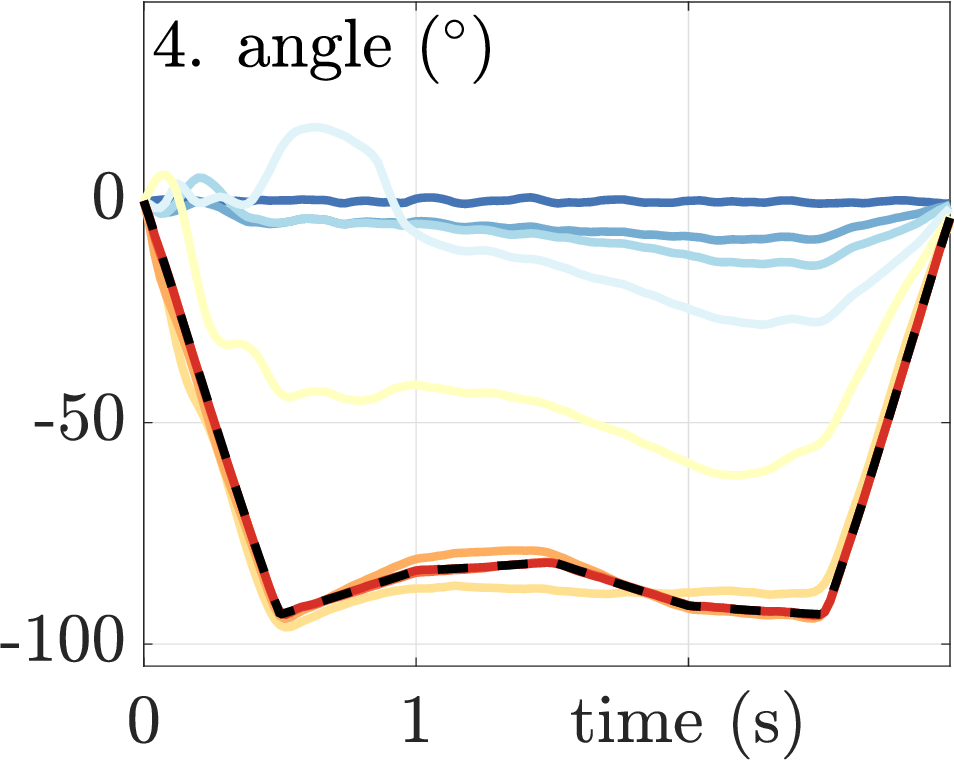}
	\includegraphics[width=0.32\linewidth]{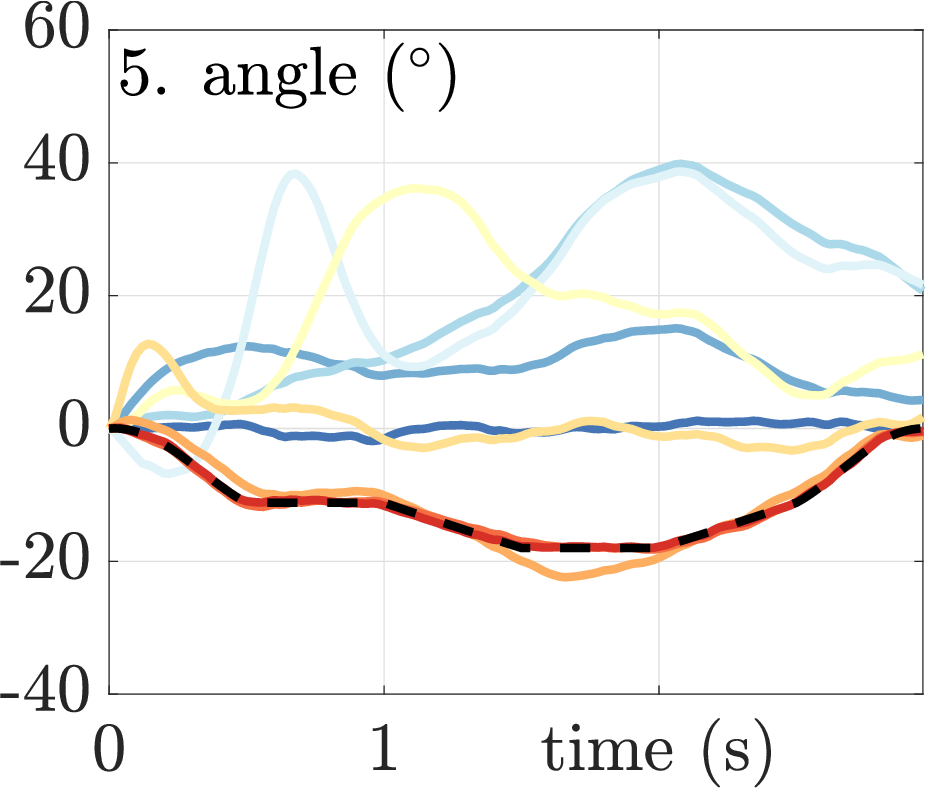}
	\includegraphics[width=0.32\linewidth]{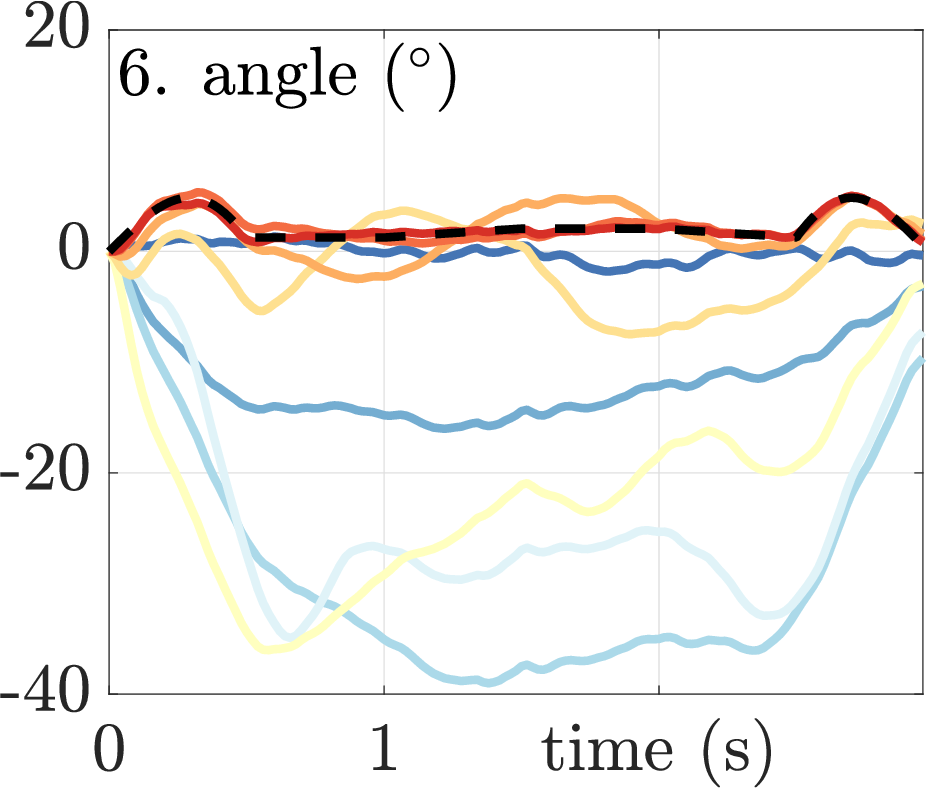}
	\hfill
	\vspace{1mm}
	\includegraphics[width=0.64\linewidth]{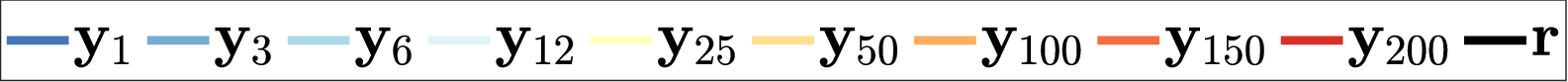}
	\caption{The simulated UR10e industrial robot is a nonlinear \ac{mimo} system with coupled dynamics. Three highly dynamic motion tasks that include discontinuities in the references are tested (top left) for different self-parametrizations. The measured joint angles in different trials (bottom) are shown for the experiment with \ac{gnodilc} and reference $\qr_1$. Despite the challenging control scenario, tracking convergence is achieved using two of three self-parametrizations without model information or manual (re-)tuning.}
	\label{fig:ur10}
\end{figure}

To investigate the applicability of the proposed \ac{dilc} to industry-typical robots with a large number of input and output variables, we consider the simulation of an UR10e robot with six degrees of freedom. 
The output variables are given by the respective joint angles, and the input variables are given by the respective motor torques.
The dynamics of the robot are simulated in \href{https://mujoco.readthedocs.io/en/stable/models.html}{MuJoCo}, and each of the torques is feedback controlled by a pre-set PID controller. 
The outputs are stored with zero-mean additive Gaussian measurement noise with a standard deviation of $10^{-5} \, \mathrm{rad}$. 
Notably, the UR10e robot exhibits nonlinear dynamics, which does not align with the initial modeling assumption of a linear time-invariant system. 

Three different references that describe various realistic motions of the robot across a wide task space are generated and depicted in Figure \ref{fig:ur10} (top left). 
Please note that the reference trajectories are highly dynamic and contain discontinuities, resulting in a challenging-to-track frequency spectrum. 
To solve the reference tracking tasks, multiple self-parametrization strategies are applied and compared. 
The initial input $\qu_0$ is drawn from a zero-mean normal distribution, and the model vector $\qm_0$ is initialized as the zero vector. 

The learning performance is depicted in Figure \ref{fig:ur10} and shows that \acs{nonodilc} and \acs{gnodilc} solve all reference tracking tasks in $100-150$ trials without knowledge about the system model or manual tuning.
In this light, the shown results follow from a \textit{plug-and-play} application of Algorithm \ref{algo:dilc}.
The progression of the output trajectories is depicted for \acs{gnodilc} and reference $\qr_1$ in the bottom plots of Figure \ref{fig:ur10}. 
Almost perfect tracking performance is achieved for all joint angles, despite the nonlinear robot dynamics and outputs of significantly different magnitudes. 
The results indicate that the method scales well to systems with a large number of inputs/outputs, as present in typical industrial robots.

Despite the promising results obtained with two self-parametrization strategies, we would like to emphasize that, in the current simulation scenario, not all self-parametrizations result in successful reference tracking. 
In Figure \ref{fig:ur10}, this can be observed for \acs{nogdilc}, which yields a stagnating error norm over the first $200$ trials. 
Presumably, this behavior can be attributed to the misalignment between the nonlinear robot dynamics and the initial assumption of a linear time-invariant system, resulting in a too-simple model representation for capturing the entire system complexity.

\subsection{Two-Link Robot}\label{sec:results_tlr}

\begin{figure}[t]
	\centering
	\includegraphics[width=0.49\linewidth]{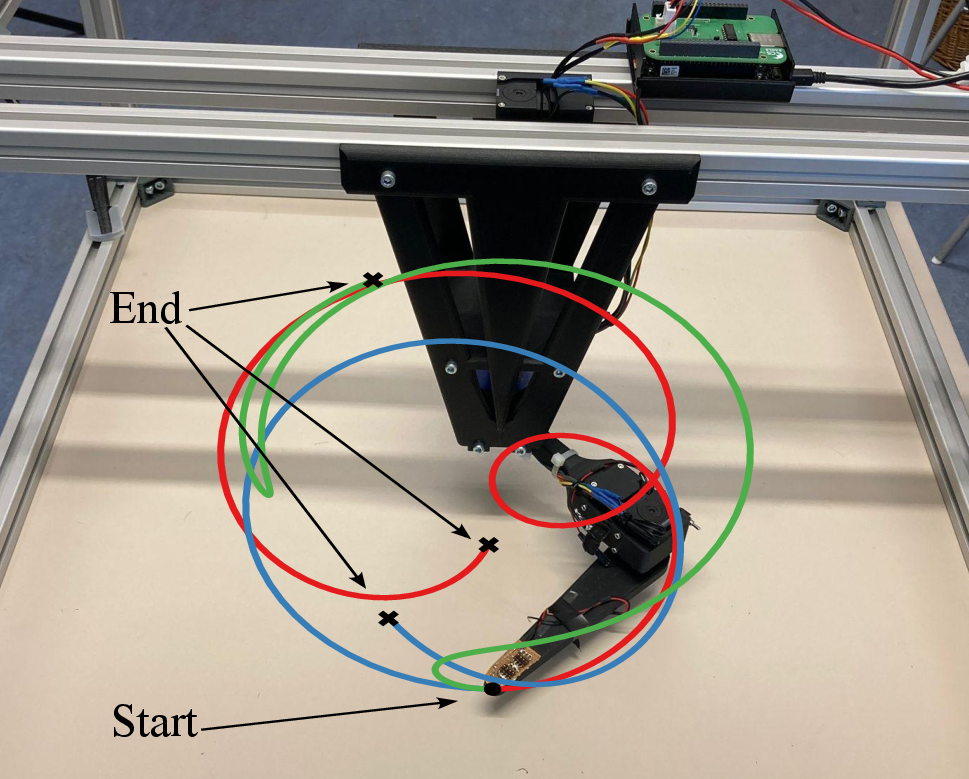}
	\includegraphics[width=0.49\linewidth]{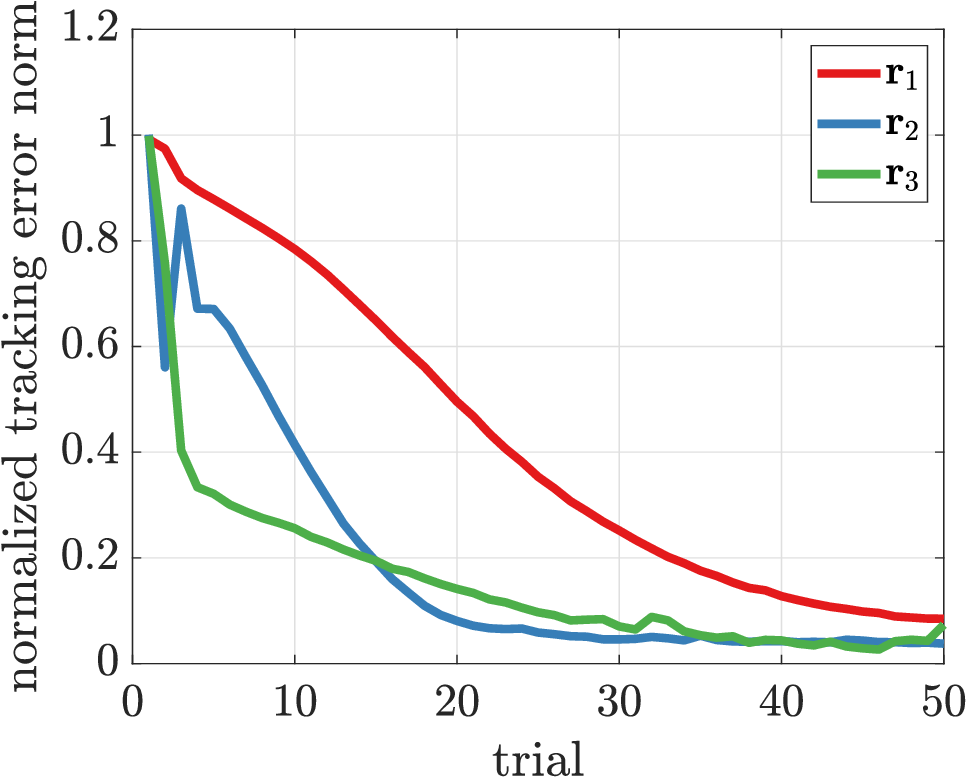}\hfill
	\vspace{2mm}
	\includegraphics[width=0.49\linewidth]{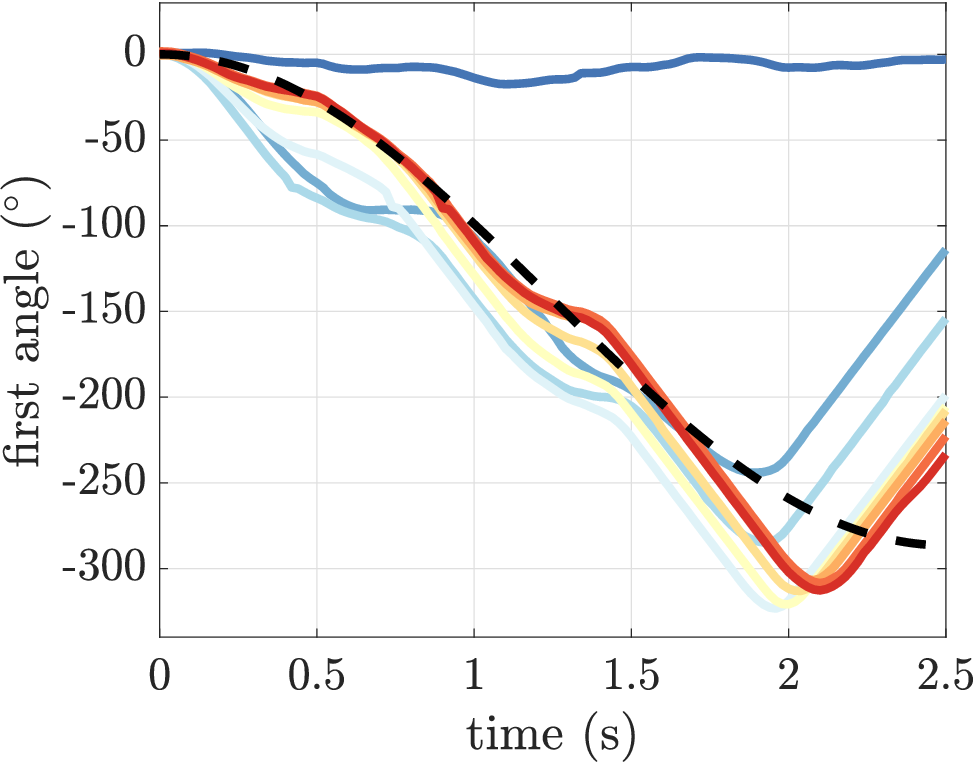}
	\includegraphics[width=0.49\linewidth]{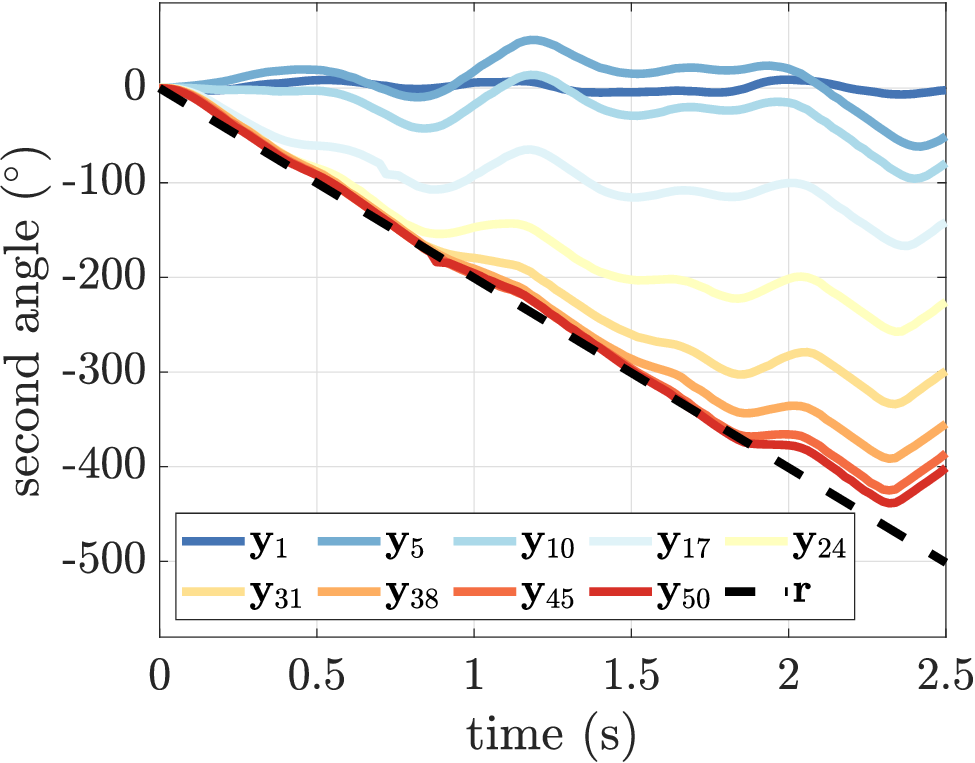}
	\caption{
		The \acl{tlr} consists of two horizontal links, each driven by a motor, and it exhibits significant gearbox backlash as well as friction effects. Three different motion tasks are tested (top left). The measured joint angles in different trials (bottom) are shown for the worst-performing reference $\qr_1$. Despite the challenging control scenario, tracking convergence is achieved in all experiments in at most $50$ trials, without model information or manually (re-)tuning the \ac{ggdilc} scheme.}
	\label{fig:twolink}
\end{figure}

In this section, the proposed \ac{dilc} framework is investigated in real-world experiments using a \ac{tlr}, see Figure \ref{fig:twolink}. 
The robot consists of two horizontal links, each driven by a motor, whose motor torques serve as input variables.
The relative angles of each of the two links are the measured output variables. 
Please note that the \ac{tlr} exhibits nonlinear dynamics, gearbox backlash, and friction effects, resulting in a significant control problem complexity.

The reference tracking tasks to be solved by the \ac{dilc} scheme are characterized by three different references, depicted in the top left plot of Figure \ref{fig:twolink}. 
As design functions, a gradient \ac{iml} and a gradient \ac{ilc} subsystem are selected, constituting a \acs{ggdilc} scheme. 

For all three references, almost monotonic learning performance is observed, and each motion task can be solved within at most $50$ trials. 
However, the learning speed varies significantly across the references, reflecting the varying complexity of the motion tasks. 
The joint space learning progress for reference $\qr_1$ is depicted in the bottom plots of Figure \ref{fig:twolink}. 
The reference-tracking performance improves gradually with the number of trials. 
In particular, reference tracking from $2\,\mathrm{s}$ seems to be difficult, which could be attributed to the gearbox backlash and friction effects in the \ac{tlr}.
The results indicate that the \ac{dilc} framework enables rapid autonomous learning in complex nonlinear real-world systems with multiple inputs and outputs, without requiring model information or manual tuning.

\subsection{Three-Wheeled Inverted Pendulum Robot}\label{sec:results_twipr}

\begin{figure}[t]
	\centering
	\includegraphics[width=0.49\linewidth]{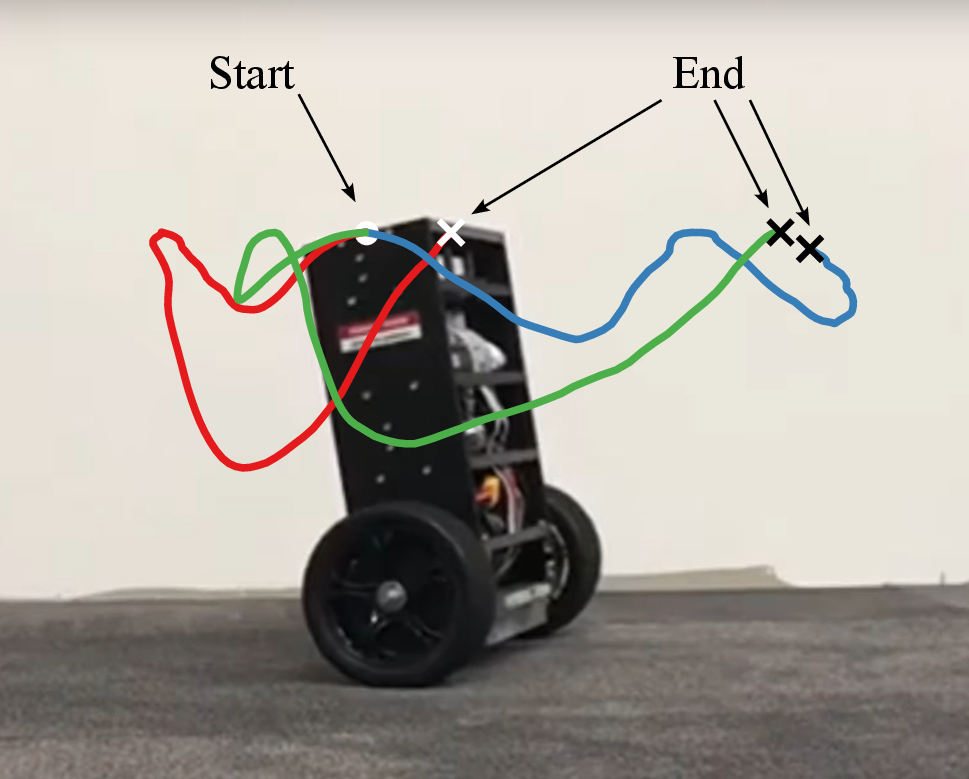}
	\includegraphics[width=0.49\linewidth]{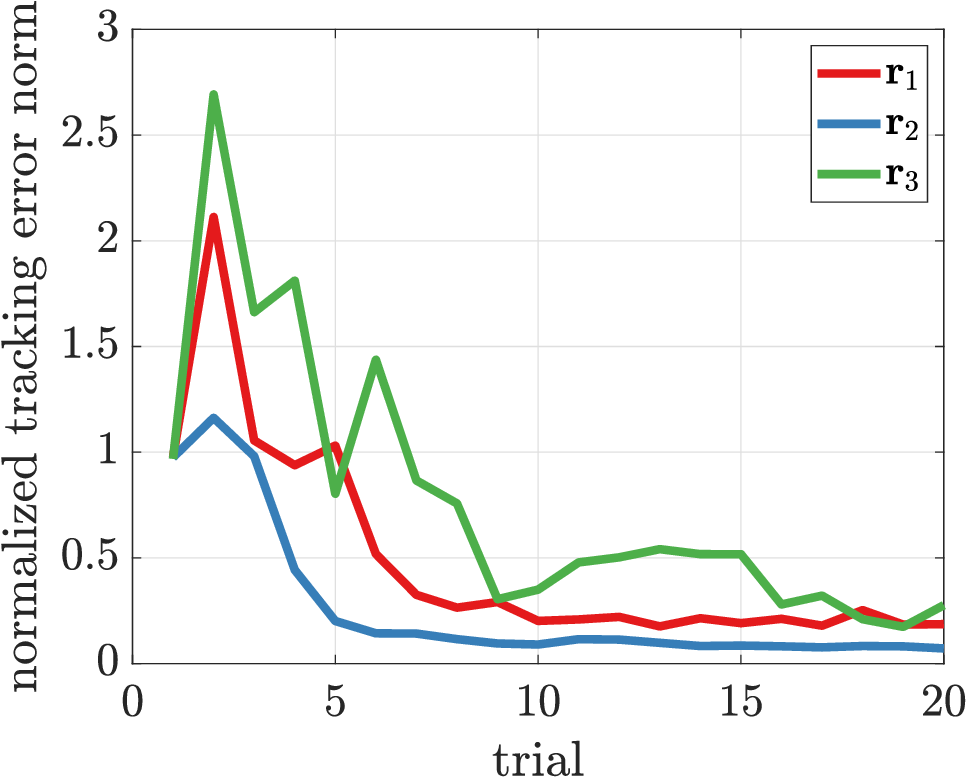}\hfill
	\vspace{2mm}
	\includegraphics[width=0.49\linewidth]{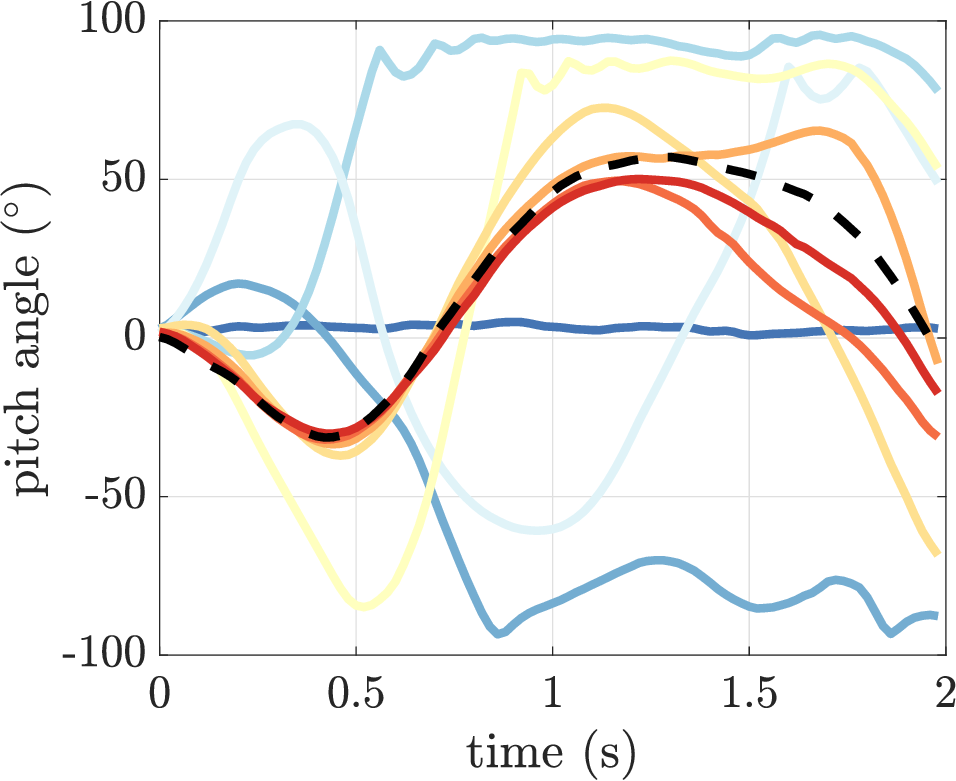}
	\includegraphics[width=0.49\linewidth]{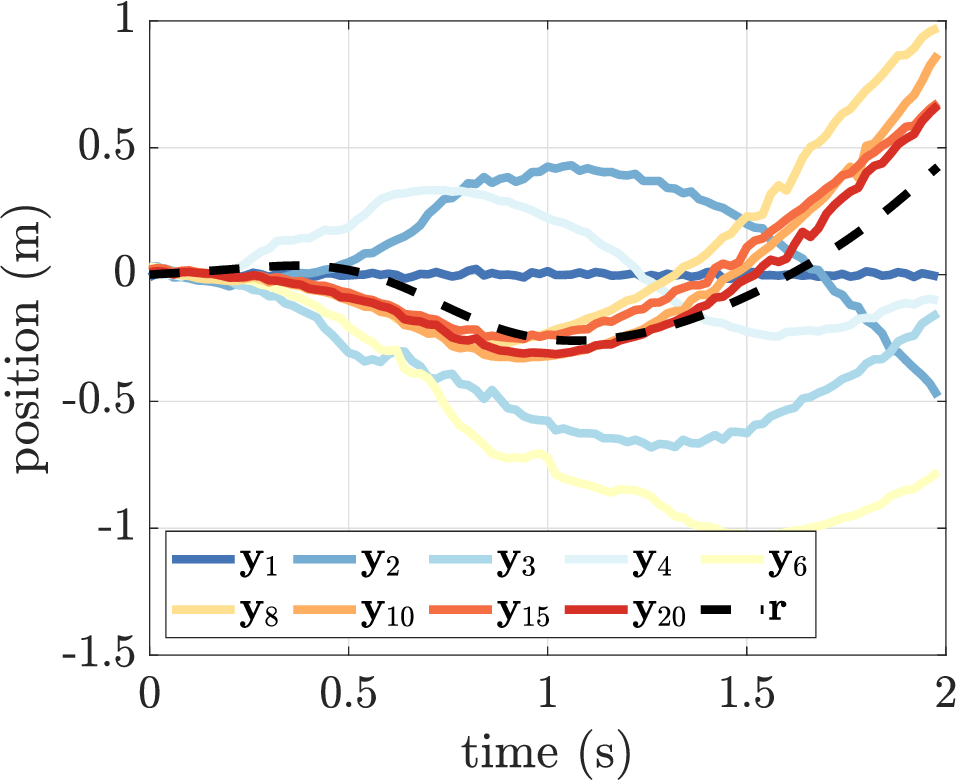}
	\caption{The \acl{twipr} (\acs{twipr}) has two motor torques (wheels and flywheel) as input variables and the robot's pitch angle and position as output variables. It exhibits strongly coupled dynamics between different inputs and outputs, as well as varying initial conditions. Three different motion tasks are tested (top left). The measured pitch angles and positions in different trials (bottom) are shown for the worst-performing reference $\qr_3$. Despite the challenging control scenario and the \ac{twipr} hitting the ground initially, tracking convergence is achieved in at most $20$ trials in all experiments, without model information or manually (re-)tuning the \ac{nogdilc} scheme.}
	\label{fig:twipr}
\end{figure}

In this section, the proposed \ac{dilc} framework is investigated in real-world experiments using a \acl{twipr} (\acs{twipr}), see Figure \ref{fig:twipr} (top left).
It consists of three major components: the robot's chassis, two tires, and a flywheel, which is mounted inside the chassis.
The robot is restricted to moving along a straight line by applying only equal torques to both tires, and the combined tire torque serves as the first input variable.
The motor torque that drives the flywheel is the second motor torque, influencing the pitch angle of the \ac{twipr}. 
The robot has two measured output variables.
The first output is the chassis' pitch angle, and the second output is the robot's longitudinal position. 
Note that the dynamics of the robots' inputs and outputs are strongly coupled and nonlinear. 
To balance the robot in the upright position, an underlying state feedback controller is employed. 
Due to the balancing mode, the initial conditions of the \ac{twipr} vary from trial to trial, which conflicts with one of the basic assumptions in \ac{ilc} to deal with repetitive settings. 

To validate the proposed \ac{dilc} framework, the robot has to solve three different reference tracking tasks, illustrated in the top left plot of Figure \ref{fig:twipr}. 
To this end, norm-optimal \ac{iml} and gradient \ac{ilc} are employed, resulting in \acs{nogdilc}, together with the self-parametrizations presented in Section \ref{sec:methods_self_parametrization}.  

The results depicted in Figure \ref{fig:twipr} show that the proposed \ac{dilc} scheme manages to reduce the tracking errors by more than $50\,\%$ within the first $10$ trials and by more than $80\,\%$ within $20$ trials. 
It can be observed that some references are significantly more challenging than others, which is reflected in initially increasing error norms. 
Potential reasons for this behavior include a significant mismatch between the learned model and the true plant dynamics, as well as physical constraints of the state space.
Specifically, the \ac{twipr} hit the ground multiple times when learning to track reference $\qr_3$, as illustrated in the bottom plots of Figure \ref{fig:twipr}. 
Nonetheless, the learning performance stabilized within a small number of trials. 

To summarize, recall that the motivation for this work is to develop a learning method that is real-world applicable, autonomous in the sense of not requiring any prior model information or manual parameter tuning, and that can deal with \ac{mimo} dynamics. 
In the presented validation, \ac{dilc} is applied to three different systems and nine different references without any prior model information or manual tuning, which underscores the scheme's autonomous learning capabilities.
Two of the experiments are conducted in real-world systems, validating real-world applicability. 
All of the considered systems are \ac{mimo}, and one system exhibits industrial-like dynamics with $6$ inputs/outputs, highlighting the method's ability to scale to large-scale systems. 
Finally, we emphasize that in each experiment, different \ac{dilc} design functions are used, showcasing the framework's capability to employ established model-based \ac{ilc} approaches for simultaneous model and control input learning without prior model information.


\section{Conclusions and Outlook}\label{sec:conclusion}
In this paper, we address the problem of autonomous learning to track references in repetitive \ac{mimo} systems. 
For these systems, we present a novel \ac{mimo} \acl{dilc} (\acs{dilc}) framework which enables simultaneous iterative learning of a control input trajectory and a system model, without requiring a prior system model or manual parameter tuning. 
We provide formal conditions for the convergence of the model and the tracking error. 
The \ac{dilc} scheme, without requiring model knowledge or manually retuning the algorithm, solves reference tracking tasks rapidly and autonomously in various high-fidelity simulations of an industrial robot and in multiple real-world \ac{mimo} systems with nonlinear dynamics. 

While the approach shows promise, we acknowledge that the learning scheme does not converge with all self-parametrizations in a nonlinear 6-axis robot simulation. 
In this light, future work should focus on extending the considered model structure in iterative model learning to linear time-varying systems, which would enable representing the dynamics of a nonlinear system along a reference trajectory.

Nonetheless, considering the rapid and \textit{autonomous} learning capabilities, we believe that \ac{dilc} has the potential to serve as an efficient building block within more complex learning frameworks for intelligent real-world systems.

\appendix

\subsection{Proof of Theorem \ref{thm:model_convergence}}\label{app:proof_thm_model_conv}
	\begin{IEEEproof}
		For the \ac{siso} case $O=1$, the proof is given in \cite[Theorem 2]{Meindl.2025}. In the following, we consider the case $O\geq 2$, following a conceptually different approach to prove the result. 

    	The proof involves three steps. 
    	First, we show that the model error norm is nonincreasing and converges to a nonnegative limit value. 
    	Second, continuity of the norm operator is exploited to relate the model error vector in the limit $j \rightarrow \infty$. 
    	Third, it is shown that, given Assumption \ref{as:excitation_condition}, the only model error vector that can persist over $O$ consecutive trials is the zero vector, implying convergence of the model error norm to $0$ and existence of a suitable $\KLfunc$-function according to Definition \ref{def:monotonic_model_convergence}.
        
		\textbf{Step 1:} Combining \eqref{eq:iml_prediction}-\eqref{eq:iml_update} yields
		\begin{equation}\label{eq:model_conv_error_dyn}
			\foralljinN \qem_{j+1} = \left(\qI - \qLh_j \qU_j \right) \qem_j \, .
		\end{equation}
		The norm of the right-hand side can be upper-bounded using the inequality, $\foralljinN $
		\begin{equation}
			\norm{ \left(\qI - \qLh_j \qU_j \right) \qem_j} \leq \norm{ \qI - \qLh_j \qU_j } \norm{\qem_j}   \, .
		\end{equation}
		Thus, a necessary condition for monotonic model convergence, corresponding to \eqref{eq:thm1_condition} in Theorem \ref{thm:model_convergence}, is
		\begin{equation}
			\foralljinN \norm{ \qI - \qLh_j \qU_j } \leq 1\, .
		\end{equation}
		For the \ac{mimo} case $O\geq2$, the matrix $\qLh_j \qU_j$ is, by construction, singular, as each of the matrices $\qLh_j \in \qRealNumbers^{O^2N \times ON}$ and $\qU_j\in \qRealNumbers^{ON \times O^2N}$ has at most rank $ON$. This implies that the norm $|| \qI - \qLh_j \qU_j ||$ cannot be less than $1$. Instead, if \eqref{eq:thm1_condition} holds, the eigenvalues of $(\qI - \qLh_j \qU_j)$ lie within the range $[-1, 1]$, which ensures a nonincreasing model error norm. Thus, in the limit $j \rightarrow \infty$, the model error norm converges to some nonnegative limit value $L_{\infty} \in [0,\infty)$, \ie
    	\begin{equation}
    		\lim_{j \rightarrow \infty} \norm{\qem_j}  = L_{\infty} \, .
    	\end{equation}
    	
    	\textbf{Step 2:} The next steps will follow a proof by contradiction. 
    	Suppose the limit value is positive, then $\qem_j$ never converges to $\qZero$. Rather, due to boundedness of the sequence, $\qem_j \rightarrow \qem_{\infty}$, and
    	\begin{equation}
    		L_{\infty} > 0 \, \Rightarrow \, \qem_{\infty} \neq \qZero \, ,
    	\end{equation} 
    	due to the continuity of the norm operator.
    	
    	\textbf{Step 3:} Consider Assumption \ref{as:excitation_condition} which provides
    	\begin{equation}
    		\foralljinN \operatorname{rank}\left(\qU_j\right) = ON \, ,
    	\end{equation}
    	implying that the null space of $\qU_j$ has dimension $(O-1) ON$. 
    	Moreover, the assumption implies, given $\qLh_j$ has full column rank, that
    	\begin{equation}
    		\foralljinN \bigcap_{i=j}^{j+O-1} \operatorname{ker}\left(\qLh_j \mathbf{U}_i\right)=\{ \qZero \} \, .
    	\end{equation}
    	Now, suppose there exists a non-zero error vector $\qem_{k}$ that persists over at least $O$ trials, \ie $	\qem_k = \qem_{k+O-1} = \mathcal{M}(\qem_{\infty}) $, where $\mathcal{M} : \qRealNumbers^{O^2 N} \rightarrow \qRealNumbers^{O^2 N}$ is the continuous map from $\qem_k$ to $\qem_{k+O-1}$, that is
    	\begin{equation}
    		\forall \qem \in \qRealNumbers^{O^2 N}, \, \mathcal{M}(\qem) \triangleq  \left( \prod_{i=j}^{j+O-1} \left(\qI - \qLh_i \qU_i \right) \right) \qem \, .
    	\end{equation}
    	Note, a stagnating model error implies that it lies in the null space of the current measurement, \ie $\qem_j \in \operatorname{ker}(\qLh_j \qU_j)$, which for the persistent error vector $\qem_k$ means
    	\begin{equation}
    		\qem_{k} \in \bigcap_{i=k}^{k+O-1} \operatorname{ker}\left(\qLh_j \mathbf{U}_i\right) \, .
    	\end{equation}
    	Therefore
    	\begin{equation}\label{eq:thm_model_error_conv_stagn_error}
    		\qem_k = \qem_{k+O-1} \, \Rightarrow \qem_k = \qZero \, .
    	\end{equation}
    	By continuity of $\mathcal{M}$ and the norm operator, \eqref{eq:thm_model_error_conv_stagn_error} is valid also for model errors $\qem_k$ close to $\qem_{\infty}$. Therefore,
    	\begin{equation}
    		\qem_{\infty} \neq \qZero \, \Rightarrow \, \norm{\mathcal{M}(\qem_{\infty})} < \norm{\qem_{\infty}} = L_{\infty} \, ,
    	\end{equation}
    	contradicting our assumption in step 2 that there exists a fixed positive limit $L_{\infty}$.
    	
    	Thus, the model error norm converges to $0$, showing the existence of a suitable $\KLfunc$-function according to Definition \ref{def:monotonic_model_convergence}.		
    \end{IEEEproof}
    
\bibliographystyle{IEEEtran} 
\bibliography{IEEEabrv,mimo_dilc}

\begin{thebibliography}{10}
\providecommand{\url}[1]{#1}
\csname url@samestyle\endcsname
\providecommand{\newblock}{\relax}
\providecommand{\bibinfo}[2]{#2}
\providecommand{\BIBentrySTDinterwordspacing}{\spaceskip=0pt\relax}
\providecommand{\BIBentryALTinterwordstretchfactor}{4}
\providecommand{\BIBentryALTinterwordspacing}{\spaceskip=\fontdimen2\font plus
\BIBentryALTinterwordstretchfactor\fontdimen3\font minus
  \fontdimen4\font\relax}
\providecommand{\BIBforeignlanguage}[2]{{%
\expandafter\ifx\csname l@#1\endcsname\relax
\typeout{** WARNING: IEEEtran.bst: No hyphenation pattern has been}%
\typeout{** loaded for the language `#1'. Using the pattern for}%
\typeout{** the default language instead.}%
\else
\language=\csname l@#1\endcsname
\fi
#2}}
\providecommand{\BIBdecl}{\relax}
\BIBdecl

\bibitem{Michalos.2016}
G.~Michalos, N.~Kousi, S.~Makris, and G.~Chryssolouris, ``{Performance
  Assessment of Production Systems with Mobile Robots},'' \emph{{Procedia
  CIRP}}, vol.~41, pp. 195--200, 2016.

\bibitem{Murphy.2004}
R.~R. Murphy, ``{Trial by Fire},'' \emph{{IEEE Robotics {\&} Automation
  Magazine}}, vol.~11, no.~3, pp. 50--61, 2004.

\bibitem{Wang.2011}
L.~Wang, E.~H.~F. {van Asseldonk}, and H.~{van der Kooij}, ``{Model Predictive
  Control-based gait pattern generation for wearable exoskeletons},''
  \emph{{IEEE International Conference on Rehabilitation Robotics}}, vol. 2011,
  p. 5975442, 2011.

\bibitem{Arimoto.1984}
S.~Arimoto, S.~Kawamura, and F.~Miyazaki, ``{Bettering operation of Robots by
  learning},'' \emph{{Journal of Robotic Systems}}, vol.~1, no.~2, pp.
  123--140, 1984.

\bibitem{Bristow.2006}
D.~A. Bristow, M.~Tharayil, and A.~G. Alleyne, ``{A survey of iterative
  learning control},'' \emph{{IEEE Control Systems Magazine}}, vol.~26, no.~3,
  pp. 96--114, 2006.

\bibitem{Wang.2009}
Y.~Wang, F.~Gao, and F.~J. Doyle, ``{Survey on iterative learning control,
  repetitive control, and run-to-run control},'' \emph{{Journal of Process
  Control}}, vol.~19, no.~10, pp. 1589--1600, 2009.

\bibitem{Freeman.2015}
C.~T. Freeman and T.~V. Dinh, ``{Experimentally verified point--to--point
  iterative learning control for highly coupled systems},''
  \emph{{International Journal of Adaptive Control and Signal Processing}},
  vol.~29, no.~3, pp. 302--324, 2015.

\bibitem{Ratcliffe.2006}
J.~D. Ratcliffe, P.~L. Lewin, E.~Rogers, J.~J. Hatonen, and D.~H. Owens,
  ``{Norm-Optimal Iterative Learning Control Applied to Gantry Robots for
  Automation Applications},'' \emph{{IEEE Transactions on Robotics}}, vol.~22,
  no.~6, pp. 1303--1307, 2006.

\bibitem{Rogers.2010}
E.~Rogers, D.~H. Owens, H.~Werner, C.~T. Freeman, P.~L. Lewin, S.~Kichhoff,
  C.~Schmidt, and G.~Lichtenberg, ``{Norm-Optimal Iterative Learning Control
  with Application to Problems in Accelerator-Based Free Electron Lasers and
  Rehabilitation Robotics},'' \emph{{European Journal of Control}}, vol.~16,
  no.~5, pp. 497--522, 2010.

\bibitem{Owens.2013}
D.~Owens, C.~Freeman, and T.~V. Dinh, ``{Norm-Optimal Iterative Learning
  Control With Intermediate Point Weighting: Theory, Algorithms, and
  Experimental Evaluation},'' \emph{{IEEE Transactions on Control Systems
  Technology}}, vol.~21, pp. 999--1007, 2013.

\bibitem{Owens.2016}
D.~H. Owens, \emph{{Iterative Learning Control: An Optimization
  Paradigm}}.\hskip 1em plus 0.5em minus 0.4em\relax London: {Springer London},
  2016.

\bibitem{Sornmo.2016}
O.~S{\"o}rnmo, B.~Bernhardsson, O.~Kr{\"o}ling, P.~Gunnarsson, and R.~Tenghamn,
  ``{Frequency-domain iterative learning control of a marine vibrator},''
  \emph{{Control Engineering Practice}}, vol.~47, pp. 70--80, 2016.

\bibitem{Helfrich.2010}
B.~E. Helfrich, C.~Lee, D.~A. Bristow, X.~H. Xiao, J.~Dong, A.~G. Alleyne,
  S.~M. Salapaka, and P.~M. Ferreira, ``{Combined $H_{\infty}$-Feedback Control
  and Iterative Learning Control Design With Application to Nanopositioning
  Systems},'' \emph{{IEEE Transactions on Control Systems Technology}},
  vol.~18, no.~2, pp. 336--351, 2010.

\bibitem{Bolder.2018}
J.~Bolder, S.~Kleinendorst, and T.~Oomen, ``{Data--driven multivariable ILC:
  enhanced performance by eliminating L and Q filters},'' \emph{{International
  Journal of Robust and Nonlinear Control}}, vol.~28, no.~12, pp. 3728--3751,
  2018.

\bibitem{Aarnoudse.2021}
L.~Aarnoudse and T.~Oomen, ``{Model-Free Learning for Massive MIMO Systems:
  Stochastic Approximation Adjoint Iterative Learning Control},'' \emph{{IEEE
  Control Systems Letters}}, vol.~5, no.~6, pp. 1946--1951, 2021.

\bibitem{Huo.2019}
B.~Huo, C.~T. Freeman, and Y.~Liu, ``{Model-free Gradient Iterative Learning
  Control for Non-linear Systems},'' \emph{{IFAC-PapersOnLine}}, vol.~52,
  no.~29, pp. 304--309, 2019.

\bibitem{Chu.2025}
B.~Chu and D.~H. Owens, \emph{{Optimal Iterative Learning Control: A
  Practitioner's Guide}}.\hskip 1em plus 0.5em minus 0.4em\relax Cham:
  {Springer Nature Switzerland}, 2025.

\bibitem{Janssens.2013}
P.~Janssens, G.~Pipeleers, and J.~Swevers, ``{A Data-Driven Constrained
  Norm-Optimal Iterative Learning Control Framework for LTI Systems},''
  \emph{{IEEE Transactions on Control Systems Technology}}, vol.~21, no.~2, pp.
  546--551, 2013.

\bibitem{Chai.2023}
S.~Chai and M.~Yu, ``{Adaptive iteration learning control with
  iteration--varying event--triggered mechanism for discrete--time nonlinear
  systems with random initial states},'' \emph{{International Journal of Robust
  and Nonlinear Control}}, vol.~33, no.~11, pp. 6135--6150, 2023.

\bibitem{Tzou.1999}
Y.-Y. Tzou, S.-L. Jung, and H.-C. Yeh, ``{Adaptive repetitive control of PWM
  inverters for very low THD AC-voltage regulation with unknown loads},''
  \emph{{IEEE Transactions on Power Electronics}}, vol.~14, no.~5, pp.
  973--981, 1999.

\bibitem{Liu.2016b}
N.~Liu and A.~Alleyne, ``{Iterative Learning Identification/Iterative Learning
  Control for Linear Time-Varying Systems},'' \emph{{Journal of Dynamic
  Systems, Measurement, and Control}}, vol. 138, no.~10, 2016.

\bibitem{Hou.2015}
Z.~Hou, Q.~Yu, J.-x. Xu, and C.~Yin, ``{A Simultaneous Iterative Learning
  Control and Dynamic Modeling Approach for A Class of Nonlinear Systems},''
  \emph{{IFAC-PapersOnLine}}, vol.~48, no.~28, pp. 368--373, 2015.

\bibitem{Rozario.2019}
R.~de~Rozario and T.~Oomen, ``{Data-driven iterative inversion-based control:
  Achieving robustness through nonlinear learning},'' \emph{{Automatica}}, vol.
  107, pp. 342--352, 2019.

\bibitem{Chi.2012}
R.~Chi, D.~Wang, Z.~Hou, and S.~Jin, ``{Data-driven optimal terminal iterative
  learning control},'' \emph{{Journal of Process Control}}, vol.~22, no.~10,
  pp. 2026--2037, 2012.

\bibitem{Chi.2015}
R.~Chi, Z.~Hou, B.~Huang, and S.~Jin, ``{A unified data-driven design framework
  of optimality-based generalized iterative learning control},''
  \emph{{Computers {\&} Chemical Engineering}}, vol.~77, pp. 10--23, 2015.

\bibitem{Chi.2017}
R.~Chi, X.~Liu, R.~Zhang, Z.~Hou, and B.~Huang, ``{Constrained data-driven
  optimal iterative learning control},'' \emph{{Journal of Process Control}},
  vol.~55, pp. 10--29, 2017.

\bibitem{Hou.2017}
Z.~Hou, R.~Chi, and H.~Gao, ``{An Overview of Dynamic-Linearization-Based
  Data-Driven Control and Applications},'' \emph{{IEEE Transactions on
  Industrial Electronics}}, vol.~64, no.~5, pp. 4076--4090, 2017.

\bibitem{Yu.2020b}
Q.~Yu, Z.~Hou, X.~Bu, and Q.~Yu, ``{RBFNN-Based Data-Driven Predictive
  Iterative Learning Control for Nonaffine Nonlinear Systems},'' \emph{{IEEE
  transactions on neural networks and learning systems}}, vol.~31, no.~4, pp.
  1170--1182, 2020.

\bibitem{Meindl.2024}
M.~Meindl, S.~Bachhuber, and T.~Seel, ``{AI-MOLE: Autonomous Iterative Motion
  Learning for unknown nonlinear dynamics with extensive experimental
  validation},'' \emph{{Control Engineering Practice}}, vol. 145, p. 105879,
  2024.

\bibitem{Meindl.2022}
M.~Meindl, D.~Lehmann, and T.~Seel, ``{Bridging Reinforcement Learning and
  Iterative Learning Control: Autonomous Motion Learning for Unknown, Nonlinear
  Dynamics},'' \emph{{Frontiers in robotics and AI}}, vol.~9, p. 793512, 2022.

\bibitem{Meindl.2024c}
M.~Meindl, R.~M{\"o}nkem{\"o}ller, and T.~Seel, ``{Autonomous Iterative Motion
  Learning (AI-MOLE) of a SCARA Robot for Automated Myocardial Injection},''
  \emph{{IFAC-PapersOnLine}}, vol.~58, no.~24, pp. 380--385, 2024.

\bibitem{Meindl.2025}
M.~Meindl, S.~Bachhuber, and T.~Seel, ``{Iterative Model Learning and Dual
  Iterative Learning Control: A Unified Framework for Data-Driven Iterative
  Learning Control},'' \emph{{IEEE Transactions on Automatic Control}}, pp.
  1--13, 2025.

\end{thebibliography}

%
%
%
%
%
%

\end{document}